\documentclass[prd,aps,showpacs,nofootinbib,nobibnotes,superscriptaddress]
              {revtex4}
\usepackage{epsfig,amssymb}
\newcommand{\bea}{\begin{eqnarray}}
\newcommand{\eea}{\end{eqnarray}}
\newcommand{\be}{\begin{equation}}
\newcommand{\ee}{\end{equation}}
\newcommand{\beast}{\begin{eqnarray*}}
\newcommand{\eeast}{\end{eqnarray*}}
\newcommand{\pkt}{\; .}
\newcommand{\kma}{\; ,}

\def\e{{\rm e}}
\newcounter{subequation}[equation]
\makeatletter \expandafter\let\expandafter\reset@font\csname
reset@font\endcsname
  {\endeqnarray\stepcounter{equation}}
\makeatother
\begin{document}

\title{CMB Temperature Anisotropy from Broken Spatial Isotropy
due to an Homogeneous Cosmological Magnetic Field}

\author{Tina Kahniashvili}
\email{tinatin@phys.ksu.edu}
\affiliation{Department of Physics,
Kansas State University, 116 Cardwell Hall, Manhattan, KS 66506, USA}
\affiliation{Department of Physics, Laurentian University,
Ramsey Lake Road, Sudbury, ON P3E 2C6, Canada}
\affiliation{National Abastumani Astrophysical Observatory, 2A
Kazbegi Ave, Tbilisi, GE-0160, Georgia}

\author{George Lavrelashvili}
\email{lavrela@itp.unibe.ch}
\affiliation{Department of Theoretical Physics, A.\ Razmadze Mathematical Institute,
1 M.\ Aleksidze,  Tbilisi, GE-0193, Georgia}

\author{Bharat Ratra}
\email{ratra@phys.ksu.edu}
\affiliation{Department of Physics, Kansas State University, 116 Cardwell Hall,
Manhattan, KS 66506, USA}

\date{July 2008}

\begin{abstract}
We derive the cosmic microwave background temperature anisotropy
two-point correlation function (including off-diagonal
correlations) from broken spatial isotropy due to an arbitrarily
oriented homogeneous cosmological magnetic field.
\end{abstract}

 \maketitle

\section{Introduction}

Improving cosmic microwave background (CMB) anisotropy
measurements are starting to make it possible to reconstruct
physical conditions  in the early Universe, and thus to constrain
modifications of the standard cosmological and particle physics
models \cite{1}. In particular, analyses of the WMAP data suggest
tentative indications of broken large-scale (statistical)
spatial  isotropy, see Refs.\ \cite{2} for early indications and
Refs.\ \cite{3} for more recent studies. Statistical large-scale
spatial isotropy is a major assumption of the standard
cosmological model and has been well tested on length scales
smaller than are probed by the large-scale CMB anisotropy data
(see Sec.\ 3 of \ Ref.\ \cite{4}). It is therefore important to
understand if the larger-scale CMB anisotropy data really
indicates that large-scale statistical spatial isotropy is broken
\cite{5}. This is part of the general program of testing for CMB
anisotropy non-gaussianity.\footnote{See the Refs.\ \cite{6} for
reviews of non-gaussian models. In the simplest inflation models,
quantum-mechanical zero-point fluctuations in a weakly coupled
scalar field during inflation provide the initial conditions
\cite{7} for a gaussian CMB anisotropy, but non-gaussian initial
conditions are possible in other inflation models.} In the last
few years there has been much discussion of the ``low'' measured
CMB temperature anisotropy quadropole moment, the asymmetry
between the CMB temperature anisotropy measured in the north and
the south, the possibility of residual systematics and foreground
emission in the data, etc. In addition to Refs.\ \cite{1,2,3,5},
for early discussions of some of these issues see Refs.\ \cite{8},
for more recent discussions see Refs.\ \cite{9}. The ``low"
measured quadropole moment was also seen in the COBE-DMR data
\cite{10}, while on smaller scales the CMB anisotropy is
consistent with gaussianity \cite{11}.

There have been several theoretical attempts to explain the CMB
temperature anisotropy large-scale anomalies as manifestations of
departure from the standard cosmological scenario, e.g.,\ via
modifications of the inflation framework, in slightly anisotropic
cosmological models, or by a preferred direction in the Universe,
etc.\ See Refs.\ \cite{cmb-anomalies-explanation} for recent
studies and Refs.\ \cite{13} for earlier works. Recently Refs.\
\cite{B} propose a cosmological magnetic field as a possible
mechanism  to explain these anomalies (the CMB temperature
anisotropy non-gaussianity that results from the magnetic field
presence has been used to limit the amplitude of such a field
\cite{cmb-anomalies-explanation-old}).

In this paper we present a formalism useful for describing CMB
temperature anisotropies in a cosmological model with a preferred
direction at the perturbation level, while the background model
preserves spatial isotropy. More specifically, we consider a
cosmological model with a uniform magnetic field pointing in a
fixed direction\footnote{Such a magnetic field can be viewed as
an approximation  of a stochastic magnetic field with
correlation length larger than the Hubble radius. A $10^{-9}$
Gauss cosmological magnetic field with correlation length larger
than the Hubble radius can be generated by quantum-mechanical
fluctuations during inflation, \cite{ratra}.}, with the
magnetic field energy density treated as a first order
perturbation, and study the CMB temperature anisotropy two-point
correlation function which reflects the magnetic-field-induced
broken spatial isotropy. A simplified version of  this problem
has been studied  in Ref.\ \cite{dky98}; here we consider an
arbitrarily oriented magnetic field. In general, a cosmological
magnetic field contributes, via the linearized Einstein
equations, to all three kinds of perturbations, scalar, vector,
and tensor, and if the amplitude of the magnetic field is large
enough ($10^{-9}$ G, or larger), there are observable
imprints on the CMB temperature anisotropies (for recent reviews
see Refs.\ \cite{mag-field-review}; for specific recent
computations see Refs.\ \cite{add}). As noted below, in our
computation we only need to consider vector perturbations.

In the model we consider here the CMB temperature two-point
correlation function reflects the presence of non-zero
off-diagonal correlations between the usual $a_{lm}$ multipole
coefficients with multipole index $l$ differing by $2$ and/or
multipole index $m$ differing by $1$ or $2$. More precisely, there
are non-zero off-diagonal correlations only for $\Delta l = \pm 2$
and $\Delta m=0$ and for $\Delta m=\pm 1$ and $\pm 2$ for both
$\Delta l =0$ and $\Delta l = \pm 2$. Some of these correlations
have been discussed in Ref.~\cite{dky98} for the case of an
homogeneous magnetic field oriented perpendicular to the
galactic  plane. Here we study the general case of an arbitrarily
oriented magnetic field, and develop a  new technique to compute
the CMB temperature anisotropy in real space. The arbitrarily
oriented magnetic field induces additional effects, not only
breaking rotational invariance breaking (resulting in non-zero
correlations between multipoles of different $l$), but also
breaks spin (parity) symmetry (resulting in non-zero off-diagonal
correlations between multipoles of different $m$). As a result,
in multipole space the $\langle a_{lm}^\star a_{l' m'} \rangle$
power spectrum is antisymmetric under exchange of $m$ and $m'$. A
similar effect occurs for Faraday rotation  of the CMB
polarization plane induced by an homogeneous magnetic field
\cite{harari}, for the cross-correlations between
$E$-polarization anisotropy and temperature or $B$-polarization
anisotropy, which vanish in the standard cosmological model in
the absence of a  primordial magnetic field.

The outline of our paper is as follows. In Sec.\ II we present
the general description of the problem, that includes a
derivation of the equations governing vorticity perturbations in
the Universe (Sec.\ II.A) and an expression for the CMB
temperature anisotropy induced by Alfv\'en waves (Sec.\ II.B). In
Sec.\ III we derive the multipole coefficient power spectrum,
which includes various $\Delta l = 0$, $\Delta l = \pm 2$,
$\Delta m = 0$, $\Delta m = \pm 1$, and $\Delta m = \pm 2$ correlations.
In Sec.\ IV we derive the real-space two-point temperature
anisotropy correlation function (the details of the computation are
summarized in App.\ B). We conclude in Sec.\ V. In App.\ A we
list useful mathematical formulae that we used in the
computations.

\section{General Description}

\subsection{Vorticity perturbations}

In this subsection we study the dynamics of  linear magnetic
vector perturbations about a spatially-flat\footnote{Current
observational data are consistent with flat spatial
hypersurfaces, see Ref.\ \cite{rv08} for a recent review.}
Friedmann-Lema\^{i}tre-Robertson-Walker (FLRW) homogeneous
cosmological spacetime background with vector metric
fluctuations. The metric tensor can be decomposed into a
spatially homogeneous background part and a perturbation part,
${g}_{\mu\nu}={\bar g}_{\mu\nu}+\delta g_{\mu\nu}$, where $\mu
,\nu \in (0, 1, 2, 3)$ are spacetime indices. For a
spatially-flat model, and working with conformal time $\eta$, the
background FLRW metric tensor ${\bar
g}_{\mu\nu}=a^2\eta_{\mu\nu}$, where
$\eta_{\mu\nu}=\mbox{diag}(-1,1,1,1)$ is the Minkowski metric
tensor and $a(\eta)$ the scale factor. Vector perturbations are
gauge dependent because the mapping of coordinates between the
perturbed physical manifold and the background is not unique.
Vector perturbations to the geometry can be described by two
three-dimensional divergence-free vector fields ${\bf A}$ and
${\bf H}$ \cite{mukhanov}, where
\begin{equation}\label{eq:S-metric-pert}
\delta g_{0i}=\delta g_{i0}=a^2 A_i,\qquad\qquad \delta
g_{ij}=a^2(H_{i,j}+H_{j,i}).
\end{equation}
Here a comma denotes the usual spatial derivative, $i, j \in (1,
2, 3)$ are spatial indices, and ${\bf A}$ and ${\bf H}$ vanish at
spatial infinity. Studying the behavior of these variables under
infinitesimal general coordinate transformations (gauge
transformations in the context of linearized gravity) one find
that ${\bf V} = {\bf A}-{\bf \dot H}$ is  gauge-invariant (the
overdot represents a derivative with respect to conformal time).
${\bf V}$ is a vector perturbation of the extrinsic curvature
\cite{bardeen}. Exploiting the gauge freedom we choose  ${\bf H}$
to be constant in time. Then the vector metric perturbation may
be described in terms of two divergenceless three-dimensional
gauge-invariant vector fields, the vector potential $\bf{V}$ and
a vector representing the transverse peculiar velocity of the
plasma, the vorticity ${\bf \Omega}={\bf v}-{\bf V}$, where
${\mathbf v}$ is the spatial part of the four-velocity
perturbation of a stationary fluid element
\cite{mkk02}.{\footnote{Given the general coordinate
transformation properties of the velocity field $ \bf v$, two
gauge-invariant quantities can be constructed, the shear ${\bf s}
= {\bf v} -{\bf {\dot H}}$ and the vorticity ${\bf \Omega}= {\bf
v } - {\bf A}$ \cite{bardeen}. In the gauge ${\bf \dot H} =0$
(i.e., ${\bf V}={\bf A}$) we get ${\bf \Omega}={\bf v}-{\bf V}$
\cite{hw97}.}} In the absence of a source the vector perturbation
${\bf V}$ decays with time (this follows from ${\bf \dot V} +
2({\dot a}/{a}) {\bf V}=0$) and so can be ignored.

 Since the fluid velocity is small
the displacement current in Amp\`ere's law may be neglected; this
implies the current $\bf J$ is determined by the magnetic field
via $ {\bf J} ={\bf\nabla} \times {\bf B }/(4\pi)$. The residual
ionization of the primordial plasma is large enough to ensure that
magnetic field lines are frozen into the plasma so the induction
law takes the form ${\dot{\bf B}} = {\bf\nabla} \times ({\bf
v}\times {\bf B})$. As a result the baryon Euler equation for
$\bf v$ has the Lorentz force ${\mathbf L({\mathbf x})}= -
{\mathbf B}({\bf x})\times\left[\nabla\times {\bf B}({\bf x})
\right]/(4\pi)$ as a source term. The photons are neutral so the
photon Euler equation does not have a Lorentz force source term.
The Euler equations for photons and baryons are
\cite{mkk02,lewis04,kr05}
\begin{eqnarray}\label{eq:V-momentum-baryon}
\dot{{\bf \Omega}}_{\gamma}+\dot{\tau}({\bf v}_{\gamma}-{\bf
v}_{b}) &=& 0,
\label{eq:V-momentum-photon}\\
\dot{{\bf \Omega}}_{b}+\frac{\dot{a}}{a}{\bf \Omega}_{b}-
\frac{\dot{\tau}}{R} ({\bf v}_{\gamma}-{\bf v}_{b}) &=&
\frac{{\bf L}\!^{(V)}\!({\mathbf x})}{a^4(\rho_b+p_b)}~,
\end{eqnarray}
where the subscripts $\gamma$ and $b$ refer to the photon and
baryon fluids, and $\rho$ and $p$ are energy density and
pressure. Here $\dot{\tau}=n_e\sigma_Ta$ is the differential
optical depth, $n_e$ is the free electron density,
$\sigma_T$  is the Thomson cross section,
$R=(\rho_b+p_b)/(\rho_\gamma+p_\gamma)\simeq3\rho_b/4\rho_\gamma$
is the momentum density ratio between baryons and photons, and
$L^{(V)}_i$ is the transverse vector (divergenceless) part of the
Lorentz force. In the tight-coupling limit
${\bf v}_{\gamma}\simeq {\bf v}_{b}$,
so we introduce the photon-baryon fluid divergenceless vorticity
${\bf \Omega}$ ($={\bf \Omega}_\gamma = {\bf \Omega}_b$)
that satisfies
%%%
\begin{equation} \label{vorticity-total}
(1+R)\dot{{\bf \Omega}}+R\frac{\dot{a}}{a}{\bf \Omega}=
\frac{{\bf L}\!^{(V)}\!({\mathbf x})}{a^4(\rho_\gamma+p_\gamma)}.
\end{equation}
%%%

As usual we consider an expansion about a spatially homogeneous
background magnetic field strength ${\bf B}_0$, writing the total
magnetic field ${\bf B}= {\bf B}_0 + {\bf B}_1$, where ${\bf
B}_1$ is a small ($|{\bf B}_1| \ll |{\bf B}_0|$) first order
inhomogeneous magnetic field strength perturbation that is
divergenceless  (${\nabla} \cdot {\bf B}_1=0$). To leading order
in ${\bf B}_1$ the induction law then gives
%%%
\begin{equation} \label{induction}
 {\dot {\bf B}}_1 = {\bf\nabla} \times {\bf v} \times {\bf B}_0~.
\end{equation}
%%%
The current in this case is determined by the magnetic field
perturbation, ${\bf J}={\bf\nabla\times} {\bf B}_1 /(4\pi)$.
Consequently the Lorentz force is ${\mathbf L({\mathbf x})}= -
{\mathbf B}_0\times\left[\nabla\times {\bf B}_1\right]/(4\pi)$.

Neglecting viscosity, which is a good approximation on scales
much larger than the Silk damping length scale, taking the time
derivative of Eq.~(\ref{vorticity-total}), for a fixed
Fourier{\footnote{For a vector field ${\bf F}$ we use
%%%
\begin{equation}
   F_j({\mathbf k}) = \int d^3\!x \,
   e^{i{\mathbf k}\cdot {\mathbf x}} F_j({\mathbf x}),~~~~~~~~~~~
   F_j({\mathbf x}) = \int {d^3\!k \over (2\pi)^3}
   e^{-i{\mathbf k}\cdot {\mathbf x}} F_j({\mathbf k}),\nonumber
\end{equation}
%%%
when Fourier transforming between real and wavenumber spaces; we
assume flat spatial hypersurfaces.}} mode $\bf k$, we get for the
transverse vorticity
%%%
\begin{equation} \label{vorticity}
{\ddot{\bf\Omega}} = \frac{ - {\bf B_0} \cdot {\bf
k}^2}{4\pi(\rho_{\gamma 0}+p_{\gamma 0})} {\bf\Omega}
\end{equation}
%%%
in the radiation dominated epoch when $R \ll 1$.  Here
$\rho_{\gamma 0}$ and $p_{\gamma 0}$ denote the present value of
the photon energy density and pressure and we have used $\dot R/R
= \dot a/a$. In general, the factor $1+R$ appearing in
Eq.~(\ref{vorticity-total}) leads to the suppression of the
vorticity amplitude due to the tight coupling between photons and
baryons, because photons being neutral are not affected by the
Lorentz force.   This suppression happens only for scales larger
than the Silk damping length scale, leaving the amplitude of
vorticity perturbations unchanged for $k>k_S$ ($k_S$ is the
wavenumber corresponding to the Silk damping length scale) \cite{silk}.

Equation (\ref{vorticity}) describes  Alfv\'en wave propagation
in the expanding Universe. These  Alfv\'en waves propagate with
phase velocity $v_A {\bf b} \cdot {\bf{\hat k}} = v_A\mu$, where
the Alfv\'en velocity $v_A=B_0/\sqrt{4\pi(\rho_{\gamma 0}
+p_{\gamma 0})}$, ${\bf b}={\bf B}_0/{B_0}$ is the unit vector in
the direction of the magnetic field, and ${\hat{\bf k}}$ is the
unit wavevector in the propagation direction. Equation
(\ref{vorticity}) has two independent solutions, conventionally
picked to be $\cos$ and $\sin$ functions. The cos solution
describes vector perturbations in the absence of the magnetic
field and thus is not of interest here. The $\sin$ solution [$
\propto \sin(v_a k\mu \eta +\phi)$, where $\phi $ is a constant
of integration]  describes  transverse Alfv\'en waves. For a
finite vorticity-energy-density, vorticity must vanish on
super-Hubble-radius scales ($k\eta \rightarrow 0$), ${\bf
\Omega}(k\eta \rightarrow 0) \rightarrow 0$, which implies $\phi
=0$, so the solution of Eq.~(\ref{vorticity}) is \cite{dky98}
\begin{equation}
 {\bf \Omega}({\bf k}, \eta) = {\bf \Omega}_0 \sin (v_A k
\eta\mu), \label{vorticity-solution1}
\end{equation}
where ${\bf \Omega}_0$ is the initial amplitude of the vorticity
perturbation in the fluid. Self-consistency{\footnote{In terms of
the magnetic field perturbation we have
\begin{equation} \label{omega-B1}
{\dot{\bf \Omega}} = \frac{i({\bf B}_0 \cdot {\bf
k})}{4\pi(\rho_{\gamma 0} + p_{\gamma 0})} {\bf B}_1 = i v_A^2 \mu
k \frac{{\bf B}_1}{|{\bf B}_0|}.\nonumber
\end{equation}
It is easy to see that ${\bf  \Omega}_0 $ is directed along ${\bf
B}_1$, and using ${\bf \Omega} = {\bf \Omega}_0 {\rm exp}(i v_A k
\mu \eta +i \phi)$, we obtain $i |{\bf \Omega}_0|v_A k \mu = i
v_A^2 \mu k |{\bf B}_1|/ |{\bf B}_0|$.}} requires
$|{\bf\Omega}_0|=|{\bf B}_1| v_A/|{\bf B_0}| =|{\bf
B}_1|/\sqrt{4\pi(\rho_{\gamma_0} +p_{\gamma_0})}$, allowing an
initial vorticity amplitude a factor $|{\bf B}_1|/|{\bf B}_0|$
($\ll 1$) smaller than the Alfv\'en velocity. Thus, Alfv\'en wave
excitations  in the Universe require (i) initial vector
(vorticity) perturbations, and (ii) a cosmological background
magnetic field. Since $v_A$ is treated as a $1/2$-order
perturbation, and the inhomogeneous magnetic field is a
first-order perturbation ($|{\bf B}_1| \ll |{\bf B}_0|$), the
amplitude of the vorticity perturbation is a 3/2-order
perturbation.

We assume that the initial vorticity perturbation spectrum in
wavenumber space is that of a stochastic gaussianly-distributed
vector field with helicity \cite{kr05},
\begin{equation}
\langle \Omega^\star_{0,i}({\mathbf k})\Omega_{0,j}({\mathbf
k'})\rangle =(2\pi)^3 \delta^{(3)} ({\mathbf k}-{\mathbf k'})
[P_{ij}({\mathbf{\hat k}}) P_{\Omega_0}(k)  + i \epsilon_{ijl}
\hat{k}_l P_{H_0}(k)]. \label{spectrum}
\end{equation}
Here $P_{ij}({\mathbf{\hat k}}) = \delta_{ij}-\hat{k}_i\hat{k}_j$
is the transverse plane projector with unit wavenumber components
$\hat{k}_i=k_i/k$, a star denotes complex conjugation,
$\epsilon_{ijl}$ is the antisymmetric tensor, and
$\delta^{(3)}({\mathbf k}-{\mathbf k'})$ is the Dirac delta
function. The power spectra $P_{\Omega_0}(k)$ and $P_{H_0}(k)$
determine the initial kinetic energy density and average helicity
of vortical motions. We approximate both spectra by simple power
laws with indices $n_\Omega$ and $n_H$.

\subsection{CMB temperature anisotropies from Alfv\'en waves}

Our aim is to study  CMB temperature anisotropies ${\Delta T}/
{T}(\mathbf{n}, \mathbf{x_0}, \eta_0)$ in the presence of an
homogeneous cosmological magnetic field $\mathbf{B}$. As usual,
$\Delta T = T(\mathbf{n}, \mathbf{x_0}, \eta_0)-\bar{T}$, where
$\bar{T}$ is the mean temperature, $\mathbf{n}$ is the unit
vector in the photon arrival direction, $\mathbf{x_0}$ is the
position of the observer, and $\eta_0$ is current conformal time
(since the big bang).

Vector perturbations induce CMB temperature anisotropies via the
Doppler and integrated Sachs-Wolfe effects \cite{dky98},
\begin{equation}
\frac{\Delta T}{T}(\eta_0,{{\mathbf n}}) = -{\mathbf
v}\cdot{{\mathbf n}}|^{\eta_0}_{\eta_{\text{dec}}} +
\int^{\eta_0}_{\eta_{\text{dec}}} d\eta\,\dot{{\mathbf
V}}\cdot{{\mathbf n}}, \label{eq:V-CMB-1}
\end{equation}
where $\eta_{\text{dec}}$ is the conformal time at decoupling.
The decaying nature of the vector potential ${\mathbf V}$ implies
that most of its contribution toward the integrated Sachs-Wolfe
term comes from near $\eta_{\text{dec}}$. Neglecting a possible
dipole contribution due to ${\mathbf v}$ today, we obtain
\cite{dky98},
%%%
\begin{equation} \label{eq:V-CMB-2}
\frac{\Delta T}{T}(\eta_0,{{\mathbf n}}) \simeq {\mathbf
v}(\eta_{\text{dec}}) \cdot{{\mathbf n}}-{\mathbf
V}(\eta_{\text{dec}})\cdot{{\mathbf n}} =  {\bf \Omega}_0 \cdot
{\bf n}
\end{equation}
%%%
(where ${\bf \Omega}_0= {\bf \Omega}(\eta_{\rm dec})$), leading
to \cite{dky98},
%%%
\begin{equation} \label{DTdef}
\frac{\Delta T}{T}(\mathbf{n,k}, \eta_0)= v_A k \eta_{\rm dec} \mu
(\mathbf{\Omega_0}(\mathbf{k}) \cdot \mathbf{n})
e^{i\mathbf{k\cdot n}\Delta \eta} \kma
\end{equation}
%%%
where wavevector ${\bf k}= k {\bf{\hat k}}$ labels the resulting
Fourier mode  after transforming from the coordinate
representation $\mathbf{x_0}$ to the momentum representation by
using  $\e^{i \mathbf{k x_0}}$, and $\Delta \eta =\eta_0-\eta_{\rm
dec} \approx \eta_0$ is the conformal time from decoupling until today.

To compute $\langle {\Delta T}/{T}(\mathbf{n}){\Delta T}/{T}
(\mathbf{n'})\rangle$ we can follow Ref.~\cite{dky98}, but the
computation is simpler if we introduce vector spherical harmonics
\cite{varshalovich89}. Using the decomposition into vector
spherical harmonics,
%%%
\begin{equation} \label{decomposition}
\mathbf{\Omega}_0(\mathbf{k})e^{i\mathbf{k\cdot n}\Delta \eta
}=\sum_{l,\lambda,m} A^{(\lambda)}_{lm} (\mathbf{k})
\mathbf{Y}^{(\lambda)}_{lm}(\mathbf{n}) \kma
\end{equation}
%%%
where  $\mathbf{Y}^{(\lambda)}_{lm}(\mathbf{n})$ (with $\lambda
=-1,0, 1$) are vector spherical harmonics (see
Eqs.~(\ref{vectorspherical}) below for definitions),  and
$A^{(\lambda)}_{lm}$ are decomposition coefficients, and taking
into account the relations $\sum_\lambda\mathbf{n \cdot
Y}^{(\lambda)}_{lm}(\mathbf{n}) = \mathbf{n \cdot
Y}^{(-1)}_{lm}(\mathbf{n}) = Y_{lm}(\mathbf{n})$ (see Eq.~(72),
p.~220, \cite{varshalovich89}, where $Y_{lm}(\mathbf{n})$ are the
usual spherical harmonics), we obtain
%%%
\begin{equation} \label{DTdef1}
\frac{\Delta T}{T}(\mathbf{n,k}, \eta_0)= v_A k \eta_{\rm dec} \mu
\sum_{l,m}A^{(-1)}_{lm} Y_{lm}(\mathbf{n}) \pkt
\end{equation}
%%%
Comparing to the conventional spherical harmonic decomposition,
${\Delta T}/{T}(\mathbf{n,k}, \eta_0)= \sum_{l,m} a_{lm}
(\mathbf{k}, \eta_0) Y_{lm}(\mathbf{n})$,  makes it possible to
relate the usual $a_{lm}$ multipole coefficients to
$A^{(-1)}_{lm}$,
%%%
\begin{equation} \label{aA}
a_{lm}(\mathbf{k}) = v_A k \eta_{\rm dec} \mu A_{lm}^{(-1)}
(\mathbf{k}) \pkt
\end{equation}
%%%
Information about the $\mathbf{\Omega}_0 (\mathbf{k})$ spectrum is
encoded in the  $A^{(-1)}_{lm}$ coefficients, which (using
Eq.~(135), p.~229, \cite{varshalovich89}) can be expressed as
%%%
\begin{equation} \label{a-1}
A^{(-1)}_{lm}({\bf k})=4\pi i^{l-1} \frac{\sqrt{l(l+1)}}{2l+1}
\bigl[j_{l-1}(k\eta_0)+j_{l+1}(k\eta_0)\bigr]
\mathbf{\Omega}_0(\mathbf{k}) \cdot {\mathbf Y}_{lm}^{(+1)
\star}(\mathbf{\hat k}) \pkt
\end{equation}
%%%
Here $j_l(x)$ are spherical Bessel functions and we have omitted
a term $\propto \mathbf{\Omega}_0 (\mathbf{k}) \cdot {\mathbf
Y}_{lm}^{(-1) \star}(\mathbf{\hat k})$ because the vorticity
vector field is transverse, $\mathbf{k}\cdot
\mathbf{\Omega}_0(\mathbf{k})=0$, and so $\mathbf{\Omega}_0
(\mathbf{k}) \cdot {\mathbf Y}_{lm}^{(-1) \star}(\mathbf{\hat
k})=0$.

We are now in a position to compute the  $\langle a^\star_{lm}
a_{l'm'} \rangle$ power spectrum,
%%%
\begin{eqnarray} \label{aspectrum}
&&\langle a^\star_{lm} a_{l'm'} \rangle = \frac{1}{(2\pi)^3} \int
dk~k^2 d\Omega_{\mathbf{\hat{k}}} a^\star_{lm} (\mathbf{k})
a_{l'm'} (\mathbf{k})~~~~ \\ \nonumber &&=\frac{2i^{l'-l}}{\pi}
\sqrt{l l'(l+1)(l'+1)} \int dk~k^2 P_{\Omega_0}(k) v_A^2
\left(\frac{\eta_{\rm
dec}}{\eta_0}\right)^2j_l(k\eta_0)j_{l'}(k\eta_0)
%\\ \nonumber &\times &
\sum_{i,j=1}^3P_{ij} \int d\Omega_{\bf \hat k} |\mu|^2
|\mathbf{Y}^{(+1) }_{lm}(\mathbf{\hat k})|^i
|\mathbf{Y}^{(+1)\star }_{l'm'}(\mathbf{\hat k})|^j \kma
\end{eqnarray}
%%%
where $d\Omega_{\bf \hat k}$ represents the solid angle volume
element, $\mu ={\bf b} \cdot {\bf\hat k}$,   and we have used
Eq.\ (\ref{spectrum}). It can be shown that initial helicity does
not contribute to the integral  in Eq.~(\ref{aspectrum}) (see
Sec.\ III  of Ref.~\cite{faraday}).  Performing the sum over $i$
and $j$  (we use Eqs.~(74), p.~220, \cite{varshalovich89}, and
vector spherical harmonics properties listed in  App.\ A.2
%\ref{vector_harmonics}
below) results in
%%%
\begin{eqnarray} \label{y7}
\langle a^\star_{lm} a_{l'm'} \rangle &=& \frac{2i^{l'-l}}{\pi}
\sqrt{ll'(l+1)(l'+1)} \int dk~k^2 P_{\Omega_0}(k) v_A^2
\left(\frac{\eta_{\rm
dec}}{\eta_0}\right)^2j_l(k\eta_0)j_{l'}(k\eta_0)
\\
&\times  &\int d{\Omega_{\bf\hat k}} \left\{
 \mathbf{Y}^{(+1)}_{lm}(\mathbf{\hat k}) \cdot
\mathbf{Y}^{(+1)\star}_{l'm'}(\mathbf{\hat k}) -(\mathbf{b} \cdot
\mathbf{Y}^{(+1)}_{lm}(\mathbf{\hat k}))
 (\mathbf{b} \cdot \mathbf{Y}^{(+1)\star}_{l'm'}(\mathbf{\hat k}))
-(\mathbf{b} \cdot \mathbf{Y}^{(0)}_{lm}(\mathbf{\hat k}))
 (\mathbf{b} \cdot \mathbf{Y}^{(0)\star}_{l'm'}(\mathbf{\hat k}))
\right\}. \nonumber
\end{eqnarray}
%%%
An advantage of this computational method over that of Ref.
\cite{dky98} is that in Eq.~(\ref{aspectrum}) we didn't need to
integrate over $d\Omega_{\mathbf n}$ and $d\Omega_{\mathbf{n'}}$.
This is similar to what happens in the total angular momentum
method \cite{hw97}.

As a consequence of the orthonormality relation,
%%%
\begin{equation} \label{y8}
\int d\Omega_{\mathbf{\hat
k}}\mathbf{Y}^{(\lambda)}_{lm}(\mathbf{\hat k}) \cdot
\mathbf{Y}^{(\lambda')\star}_{l'm'}(\mathbf{\hat k})=
\delta_{\lambda\lambda^\prime}\delta_{ll'}\delta_{mm'},
\end{equation}
%%%
the first term in the $d\Omega_{\bf\hat k}$ integral in Eq.\
(\ref{y7}) results in the usual diagonal correlations. The second
term in Eq.\ (\ref{y7}) includes non-zero correlations for  $l=l'$
and $l=l'\pm
 2$, as well as $m=m'$, $m= m'\pm 1$, and  $m=m'
 \pm 2$ (if $\mathbf{b||z}$ there are non-zero correlations only
for $m=m'$ \cite{dky98}), while the third term includes non-zero
correlations for $l=l'$ and  $m=m'$, $m= m'\pm 1$, and $m=m'
 \pm 2$ (again, if $\mathbf{b||z}$ there are non-zero correlations
 only for $m=m'$ \cite{dky98}).

To simplify the computation, we rewrite the last two terms in the
$d\Omega_{\bf\hat k}$ integral in Eq.~(\ref{y7}) in terms of
Wigner $D $ functions.  Wigner $D $ functions relate helicity
basis vectors $ \mathbf{e'}_{\!\pm 1}=\mp (\mathbf{e}_\Theta
 \pm i\mathbf{e}_\phi)/\sqrt{2}$ and $\mathbf{e'}_{\! 0}=\mathbf{e}_r$
to  spherical basis vectors $\mathbf{e}_{\pm 1}=\mp (\mathbf{e}_x
\pm i\mathbf{e}_y)/\sqrt{2}$ and $\mathbf{e}_0=\mathbf{e}_z$ (see
Eq.~(53), p.~11,  \cite{varshalovich89}) through
%%%
\begin{eqnarray} \label{basis4}
\mathbf{e'}_{\!\!\mu}= \sum_\nu D^1_{\nu \mu} (\phi, \Theta, 0)
\mathbf{e}_\nu, ~~~~~\nu, \mu=-1,0,1 \pkt
\end{eqnarray}
%%%
In both  the spherical basis and the helicity basis the following
relations hold: $\mathbf{e}_\nu \mathbf{e}^\mu=\delta_{\nu \mu}$,
$\mathbf{e}^\mu =(-1)^\mu \mathbf{e}_{-\mu}$,
$\mathbf{e}^\mu=\mathbf{e}^\star_\mu$, $\mathbf{e}_\mu \times
\mathbf{e}_{\nu}=-i\epsilon_{\mu \nu \lambda} \mathbf{e}_\lambda$.

Vector spherical harmonics may be expressed in terms of  Wigner
$D$ functions in the helicity basis where the angles $\Theta$ and
$\phi$ are defined in terms of the unit wavevector $\mathbf{\hat
k}$, see Eqs.~(\ref{y9}). Using these relations the last two
terms in the $d\Omega_{\bf\hat k}$  integral in Eq.~(\ref{y7})
become
\begin{eqnarray}
\frac{-1}{4\pi}\sqrt{(2l+1)(2l'+1)}&& \left[ (\mathbf{b} \cdot
\mathbf{e'}_{\!\!+1}(\Theta, \phi)) (\mathbf{b} \cdot
\mathbf{e'}_{\!\!+1}(\Theta,\phi))^\star
D^{l}_{-1,-m}(0,\Theta,\phi)
D^{l'\star}_{-1,-m'}(0,\Theta,\phi)   \right. \nonumber\\&&
\left.~+(\mathbf{b} \cdot \mathbf{e'}_{\!\!-1}(\Theta, \phi))
(\mathbf{b} \cdot \mathbf{e'}_{\!\!-1}(\Theta,\phi))^\star
D^{l}_{1,-m}(0,\Theta,\phi) D^{l'\star}_{1,-m'}(0,\Theta,\phi)
\right]. \label{y10}
\end{eqnarray}
The unit vector field ${\bf b}$ may be written in terms of
spherical harmonics (see Eq.~(13), p.~13, \cite{varshalovich89}),
and using Eqs.~(\ref{basis4}), (\ref{y10}), and (\ref{w}), we
obtain for the $d\Omega_{\bf\hat k}$ integral in  Eq.~(\ref{y7}),
%%%
\begin{eqnarray} \label{y11}
&&\int d{\Omega_{\bf\hat k}} \left\{
 \mathbf{Y}^{(+1)}_{lm}(\mathbf{\hat k}) \cdot
\mathbf{Y}^{(+1)\star}_{l'm'}(\mathbf{\hat k}) -(\mathbf{b} \cdot
\mathbf{Y}^{(+1)}_{lm}(\mathbf{\hat k}))
 (\mathbf{b} \cdot \mathbf{Y}^{(+1)\star}_{l'm'}(\mathbf{\hat k}))
-(\mathbf{b} \cdot \mathbf{Y}^{(0)}_{lm}(\mathbf{\hat k}))
 (\mathbf{b} \cdot \mathbf{Y}^{(0)\star}_{l'm'}(\mathbf{\hat k}))
\right\}  \\ \nonumber &&~~~=
\delta_{ll'}\delta_{mm'}-\frac{2\pi}{3}
\bigl\{1+(-1)^{l+l'}\bigr\} \sqrt{(2l+1)(2l'+1)} \int_{0}^\pi\!\!
d\Theta~ {\rm sin} \Theta \sum_{\nu, \nu' =-1}^1 (-1)^{\nu+\nu'}
\delta_{m,m'-\nu+\nu^\prime} Y^\star_{1, \nu}(\mathbf{b}) Y_{1,
\nu'}(\mathbf{b}) \\ \nonumber &&~~~~~~~~~~~~~~~~~~~~\times
d^1_{-\nu, 1}(\Theta) d^1_{-\nu^\prime, 1} (\Theta)
d^l_{-1,-m}(\Theta) d^{l'}_{-1,-m'}(\Theta) \pkt
\end{eqnarray}
%%%
Here the $ d^{l}_{mm'}(\beta)$ functions are defined in Sec.~4.3
of Ref.~\cite{varshalovich89}, and we have used the reality of
these functions as well as the relations
$d_{m,m'}^l(\pi-\Theta)=(-1)^{l-m'} d^l_{-m, m'}(\Theta)=
(-1)^{l+m} d^l_{m, -m'}(\Theta)$ (Eq.~(1), p.~79,
\cite{varshalovich89}). The expression in Eq.~(\ref{y11})
indicates that there are in general nonzero correlations for
$l=l'\pm a$, where $a$ is even. In addition there are the
following possibilities: (i) when $\nu=\nu^\prime$ there are
nonzero correlations for $m=m'$; (ii) when $|\nu -\nu^\prime|=1$
there are nonzero correlations for $m=m'\pm 1$; and (iii) when
$|\nu-\nu^\prime|=2$ there are nonzero correlations for $m=m'\pm
2$.

It is convenient to introduce the notation
%%%
\begin{equation} \label{int1}
I_d^{(l,l')}=\frac{2}{\pi}\int d k~k^2 P_{\Omega_0}(k) v_A^2
\left(\frac{\eta_{\rm dec}}{\eta_0}\right)^2
j_l(k\eta_0)j_{l'}(k\eta_0) \pkt
\end{equation}
%%%
Then Eq.~(\ref{y11}), for the multipole coefficients power
spectrum, may be rewritten as
%%%
\begin{eqnarray} \label{y15}
\langle a_{lm}^\star a_{l'm'} \rangle &=& {i^{l-l'}}
\sqrt{ll'(l+1)(l'+1)}I_d^{(l,l')}  \\ \nonumber &\times& \left[
\delta_{ll'}\delta_{mm'} -2\pi \{1+(-1)^{l+l'}\}
\sqrt{(2l+1)(2l'+1)} \int_0^\pi d\Theta \sin\Theta S_{mm'}
(\Theta, \Theta_B, \phi_B) d^l_{-1,-m}(\Theta)
d^{l'}_{-1,-m'}(\Theta)\right] \kma
\end{eqnarray}
%%%
where we have defined
%%%
\begin{equation} \label{smm}
S_{mm'}(\Theta, \Theta_B, \phi_B) = \frac{1}{3} \sum_{\nu, \nu' =
-1}^1 (-1)^{\nu+\nu^\prime}Y^\star_{1, \nu}(\mathbf{b})
Y_{1,\nu^\prime}(\mathbf{b}) \delta_{m,m'-\nu+\nu^\prime}
d^1_{-\nu, 1} (\Theta) d^1_{-\nu^\prime, 1} (\Theta) \pkt
\end{equation}
%%%

For $l+l^\prime$ odd the $1+(-1)^{l+l'}$ factor in Eq.~(\ref{y15})
is zero and so is the off-diagonal piece. For $l+l^\prime$ even we
must sum over $\nu$ and $\nu^\prime$ in Eq.~(\ref{smm}). Using
expressions for $d^l_{\nu, 1}(\Theta)$ (Eq.~(16), p.~78,
\cite{varshalovich89}) and $Y_{1,\nu}(\mathbf{b})$ (Eq.~(2),
p.~155, \cite{varshalovich89}), the double summation results in 5
different terms (corresponding to nonzero correlations for
$m=m'$, $m=m'\pm 1$, and $m=m'\pm 2$),
%%%
\begin{eqnarray} \label{y13}
S_{mm'}(\Theta, \Theta_B, \phi_B) = \frac{1}{16\pi} &\Bigl\{
&\!\!\!\! [(1+\cos^2\Theta_B)- (3\cos^2\Theta_B-1) \cos^2\Theta ]
\delta_{mm'} \Bigr.
\\ \nonumber
&&~ -2\sin\Theta_B \cos\Theta_B \sin\Theta \cos\Theta
[e^{-i\phi_B} \delta_{m,m'-1}+ e^{i\phi_B} \delta_{m,m'+1}]
\left.\right. \nonumber\\&&~-\left.\frac{1}{2}\sin^2\Theta_B
\sin^2\Theta [e^{-2i\phi_B} \delta_{m,m'-2}+ e^{2i\phi_B}
\delta_{m,m'+2}] \right\}.\nonumber
\end{eqnarray}
%%%
When $\mathbf{b}||\mathbf{z}$, $\sin\Theta_B = \phi_B=0$ and only
one term survives,
%%%
\begin{eqnarray} \label{y14}
S_{mm'}(\Theta, 0, 0) = \frac{1}{3} \sum_{\nu,\nu'=-1}^1
Y^\star_{1, \nu}(\mathbf{b}) Y_{1,\nu'}(\mathbf{b})
\delta_{m,m'-\nu+\nu^\prime} d^1_{-\nu,
1}(\Theta)d^1_{-\nu^\prime, 1}(\Theta)=
\frac{1}{8\pi}(1-\cos^2\Theta) \delta_{mm'} \pkt
\end{eqnarray}
%%%
This reproduces the result of Ref.~\cite{dky98}.

\section{Multipole coefficient power spectrum}

For an arbitrary $\mathbf{B}_0$ the multipole coefficient  power
spectrum is a function of two spherical angles, corresponding to
 the angular separation between ${\bf b}$ and directional vectors
${\bf n}$ and ${\bf n'}$.  The amplitude of the power spectrum
depends on $v_A$, $P_{\Omega_0}$, and the photon travel distances
from decoupling until today. In this Section we study diagonal
(in terms of $l$) $l=l'$ and off-diagonal $l=l'\pm 2$
correlations separately. We note that the terms with $l=l'$ and
$m=m'$ we compute here are purely due to the presence of the
magnetic field and must be added to the usual CMB temperature
anisotropy terms induced by other sources (for example, scalar
and/or tensor perturbations generated by quantum fluctuations
during inflation). Since the magnetic field amplitude is small we
ignore correlations between magnetic field and scalar (or other)
perturbations.

\subsection{$l=l'$ correlations}

For $l=l'$, the integral expression of Eq. (\ref{int1}) takes the
form,
%%%
\begin{equation} \label{int11}
I_d^{(l,l)}=\frac{2}{\pi}\int d k~ k^2 P_{\Omega_0}(k) v_A^2
\left(\frac{\eta_{\rm dec}}{\eta_0}\right)^2 j_l^2(k\eta_0) \pkt
\end{equation}
%%%
The corresponding  multipole coefficients power spectrum, Eq.
(\ref{y15}), becomes,
%%%
\begin{eqnarray}
C_l^{(m, m')}(\Theta_B, \phi_B) &= & \langle a_{lm}^\star a_{l m'}\rangle
\nonumber
\\
&=& \Bigl\{ (3\cos^2\Theta_B -1)I_{0,0}\delta_{mm'} +2\sin\Theta_B \cos\Theta_B
\Bigl[ e^{-i\phi_B}
\delta_{m, m'-1} I_{0,-1} + e^{i\phi_B} \delta_{m, m'+1}
I_{0,+1} \Bigr] \Bigr.
\nonumber
\\
&& \left.~~ +\frac{1}{2}\sin^2\Theta_B
\Bigl[ e^{-2i\phi_B} \delta_{m, m'-2} I_{0,-2} +  e^{2i\phi_B}
\delta_{m, m'+2} I_{0, +2} \Bigr] \right \}  I_d^{(l,l)},  \label{30}
\end{eqnarray}
%%%
and for $\mathbf{b||z}$, $\sin\Theta_B=0$ and it is easy to
recover the result of Ref. \cite{dky98}, $C_l(m) =
(3\cos^2\Theta_B-1)I_{0,0}I_d^{(l,l)}$. In Eq. (\ref{30}) the
coefficient $I_{0, 0}$ is
\begin{eqnarray}
I_{0,0}(l,m,\Theta_B)=\frac{l(l+1)}{(2l-1)(2l+3)} \left\{
\frac{l(l+1)+(l^2+l-3)\cos^2\Theta_B}{3\cos^2\Theta_B -1} - m^2
\left[1-\frac{3}{l(l+1)}\right] \right\}, \label{I00}
\end{eqnarray}
and
\begin{eqnarray}
I_{0, \pm 1}(l, m)&=& - \frac{l^2+l-3}{(2l-1)(2l+3)} \left (m \mp
\frac{1}{2}\right)\sqrt{(l\pm m)(l \mp m+1)},\label{I01}
\\
I_{0, \pm 2}(l,m) &= &- \frac{l^2+l-3}{(2l-1)(2l+3)} \sqrt{(l\pm
m)(l\pm m-1)(l \mp m+1)(l \mp m+2)}.
\end{eqnarray}
For an arbitrarily oriented magnetic field, even when $l=l'$,
non-zero $I_{0,\pm 1}$ and  $I_{0, \pm 2}$ indicate that there
are non-zero non-equal $m, m'$ correlations. The coefficients
$I_{0, \pm a}(m)$ have the following symmetries,
%%%
\begin{equation}
I_{0, \pm a}(m)= (-1)^a I_{0, \mp a}(-m) = I_{0, \mp a}(m \mp a)
= I_{0, \pm a}(-(m \pm a)) \kma \label{Isymmetry}
\end{equation}
%%%
where $a=|m-m'|$ and thus takes values $0, 1$, and $2$.

%\subsubsection{Magnetic field direction testing}
Taking the complex conjugate of $C_l^{(m,m')}$ it is
straightforward  to see that
%%%
\begin{equation} \label{mm'3}
C_l^{(m, m')\star}(\Theta_B, \phi_B) = C_l^{(m', m)} (\Theta_B,
\phi_B)=C_l^{(m, m')} (\Theta_B, -\phi_B) \kma
\end{equation}
%%%
so exchanging $m$ and $ m'$ corresponds to replacing  $\phi_B$
by  $-\phi_B$, and effectively corresponds to complex conjugation.

The imaginary part of $C_l^{(m,m')}$ is
\begin{eqnarray}
A_{m,m'}^{l=l'}\!(\Theta_B, \phi_B) &=& -\frac{i}{2}\left\{
C_l^{(m,m')}- C_l^{(m,m') \star}\right\}=
{\mbox{Im}}\Bigl(\langle a_{l,m}^\star a_{l, m'}\rangle \Bigr)
\nonumber
\\
&= & -\sin\Theta_B \sin \phi_B \Bigl\{
2\cos\Theta_B \Bigl[ \delta_{m,m'-1} I_{0,
-1}(m) - \delta_{m,m'+1} I_{0, +1}(m) \Bigr] \Bigr.
\nonumber
\\
\Bigl. &&~+ \sin\Theta_B
\cos\phi_B \Bigl[\delta_{m,m'-2}I_{0, -2}(m) - \delta_{m,m'+2}
I_{0, +2}(m) \Bigr]\Bigr\}  I_d^{l,l}. \label{CLIM}
\end{eqnarray}
So a measured imaginary part of $\langle a_{l,m}^\star a_{l,
m'}\rangle$  will indicate the direction of the magnetic field in
space. For a magnetic field along ${\bf z}$ the imaginary part
vanishes. The imaginary part also vanishes when $\phi_B=0$. These
imply that $A_{m,m'}^{l=l'}(\Theta_B, \phi_B) \propto |{\bf b}
\times {\bf \hat z}|$.

\subsection{$l=l'\pm 2$ correlations}

Making use of the symmetries, we need to determine
%%%
\begin{equation} \label{DL}
\langle a_{l-1,m}^\star a_{l+1,m'} \rangle =
D_{l-1,l+1}^{(m,m')}(\Theta_B,\phi_B), ~~~~~~~~~~~ \langle
a_{l+1,m}^\star a_{l-1,m'} \rangle =
D_{l+1,l-1}^{(m,m')}(\Theta_B,\phi_B).
\end{equation}
%%%
Proceeding in a similar way as for the $l=l'$ case  we find
%%%
\begin{eqnarray} \label{1}
D_{l \mp 1, l \pm 1}^{(m,m')}(\Theta_B,\phi_B)&=&
\Bigl\{(3\cos^2\Theta_B - 1)
\delta_{mm'} I_{\pm 2, 0}  \Bigr. \nonumber \\
&&~~ +2\sin\Theta_B \cos\Theta_B \left[ e^{-i\phi_B} \delta_{m,
m'-1} I_{\pm 2, -1} +
e^{i\phi_B} \delta_{m, m'+1} I_{\pm 2, +1}\right] \nonumber \\
&& ~~+ \frac{1}{2}\sin^2\Theta_B \left[e^{-2i\phi_B} \delta_{m,
m'-2} I_{\pm 2, -2} + e^{2i\phi_B} \delta_{m, m'+2} I_{\pm 2,
+2}\right ] \Bigr\}  I_d^{(l\mp 1, l \pm 1)}.
\end{eqnarray}
%%%
%For $\Theta_B=0$ what we assumed in DKY, we could easy got the result of DKY.

Defining the coefficient
%%%
\begin{equation} \label{coeff}
\mathcal{I}(l) = \frac{(l+2)(l-1)}{2(2l+1)\sqrt{(2l-1)(2l+3)} },
\end{equation}
%%%
we list and discuss separately the $m=m'$ ($I_{\pm 2, 0}$),
$m=m'\pm 1$ ($I_{\pm 2, \pm 1}$), and $m = m'\pm 2$ ($I_{\pm 2,
\pm 2}$) term coefficients of Eq.\ (\ref{1}) in what follows.

%\subsubsection{$m=m'$ term, - DKY}

For the $m=m'$ term we find
%%%
\begin{equation} \label{I20}
I_{\pm 2, 0} (l,m)=- \sqrt{(l+m)(l-m)(l-m+1)(l+m+1)}~\mathcal{I},
\end{equation}
%%%
which results in
%%%
\begin{equation} \label{DKY-DL}
D_{l-1,l+1}^{(m=m')} = \langle a_{l-1,m}^\star a_{l+1,m} \rangle
=\langle a_{l+1,m}^\star a_{l-1,m} \rangle = D_{l+1,l-1}^{(m=m')}
=(3\cos^2\Theta_B - 1)I_{+2,0} I_d^{(l-1, l+1)},
\end{equation}
%%%
where $I_d^{(l,l')}$ is defined in Eq.~(\ref{int1}). From Eq.\
(\ref{DKY-DL}), when $\Theta_B=0$ the part of $\langle a_{lm}^\star
a_{l'm'}\rangle$ proportional to $\delta_{mm'}$  is
$2I_{+2,0}I_d^{(l-1,l+1)}$, which coincides with the result of
Ref. \cite{dky98}.

%\subsubsection{$m=m' \pm 1$ term}

The part of the right hand side of Eq. (\ref{1}) proportional to $
\sin\Theta_B \cos\Theta_B$ contains terms  proportional to
$\delta_{l\mp 1, l\pm 1}$ and $\delta_{m,m'-1}$ or
$\delta_{m,m'+1}$. The coefficients in these terms are
\begin{equation}
I_{+2, \mp 1}(l,m)=\sqrt{(l \mp m)(l \pm m)(l \pm m+1)(l \pm
m+2)}~\mathcal{I}= \sqrt{\frac{(l \mp m)!}{(l \mp m-1)!}\frac{(l
\pm m+2)!}{(l \pm m-1)!}} ~\mathcal{I}, \label{I2-}
\end{equation}
and
\begin{equation}
I_{-2, \mp 1}(l,m)=-\sqrt{(l \pm m+1)(l \mp m+1)(l \mp m)(l \mp
m-1)} ~\mathcal{I} =-\sqrt{\frac{(l \pm m+1)!}{(l \pm
m)!}\frac{(l \mp m+1)!}{(l \mp m-2)!}} ~\mathcal{I}.  \label{I2+}
\end{equation}
These have the following symmetries,
\begin{equation}
I_{\pm 2, \pm 1}(m)=-I_{\pm 2, \mp 1}(-m)= I_{\mp 2, \mp 1}(m \mp
1) =-I_{\mp 2, \pm 1}(-(m \mp 1)),  \label{I3ll'pm22}
\end{equation}
i.e., the cross correlations between $l-1$ and $ l+1$ multipole
coefficients  are the negative of those between $l+1$ and $l-1$
multipole coefficients provided $m$ is replaced by  $-(m \pm 1)$.

%\subsubsection{$m=m'\pm 2$ term}

The coefficients of the last set of terms in Eq. (\ref{1}) with
$m=m'\pm 2$ are
\begin{equation} \label{I4-}
I_{+2, \mp 2}(l,m)=-\sqrt{(l \pm m)(l \pm m+1)(l \pm m+2)(l \pm
m+3)}~\mathcal{I} =-\sqrt{\frac{(l \pm m+3)!}{(l \pm m-1)!}}
~\mathcal{I},
\end{equation}
%%%
and
%%%
\begin{equation} \label{I4+}
I_{-2, \mp 2 }(l,m)=-\sqrt{(l \mp m+1)(l \mp m)(l \mp m-1)(l \mp
m-2)} ~\mathcal{I} =-\sqrt{\frac{(l \mp m+1)!}{(l \mp m-3)!}}
~\mathcal{I}.
\end{equation}
%%%
These have the following symmetries,
\begin{equation}
I_{\pm 2, \pm 2}(m)= I_{\pm 2, \mp 2}(-m)=I_{\mp 2, \mp 2}(m \mp
2)=I_{\mp 2, \pm 2}(-(m \mp 2)), \label{I5ll'pm22}
\end{equation}
i.e., the correlations between $l-1$ and $l+1$ multipole
coefficients are the negative of those between  $l+1$ and $l-1$
multipole coefficients  provided $m$ is replaced by  $-(m \pm 2)$.

Equations (\ref{I20}), (\ref{I3ll'pm22}), and (\ref{I5ll'pm22})
can be combined into one set of equations that reflect the
symmetry of the $I_{\pm 2, \pm a}$ ($a=0, 1, 2)$ coefficients,
similar to Eq. (\ref{Isymmetry}), resulting in one set of
equations for both the $l=l'$ and $l=l'\pm 2$ cases,
\begin{equation}
I_{\pm b, \pm a}(m)= (-1)^a I_{\pm b, \mp a}(-m)=I_{\mp b, \mp
a}(m \mp a)=I_{\mp b, \pm a}(-(m \mp a)), \label{Iabsymmetry}
\end{equation}
where $b=0$ or $2$. On the other hand, the magnitude of the
cross-correlation coefficients for $l=l'$ and $l = l'\pm 2$ are
different; while  all terms for the $l=l'\pm 2$ case are
proportional to ${\mathcal I}$, this is not true  for the $l=l'$
coefficients.

%\subsubsection{Imaginary part of  $l=l'\pm 2$ cross-correlations multipole power spectra}

The  non-zero off-diagonal correlations power spectrum terms
$D^{m,m'}_{l \mp 1, l\pm 1}(\Theta_B,\phi_B)$ are given by Eq.
(\ref{1}). Taking the  complex conjugate we see
\begin{equation}
D_{l \mp 1, l \pm 1}^{(m, m')\star}(\Theta_B, \phi_B) = D_{l \mp
1, l \pm 1}^{(m', m)} (\Theta_B, \phi_B)=D_{l \mp 1, l \pm
1}^{(m, m')} (\Theta_B, -\phi_B), \label{mm'4}
\end{equation}
so, as for the $C_l^{(m,m')}$ function in Eq.~(\ref{mm'3}), complex
conjugation is equivalent to exchanging $\phi_B$ and  $-\phi_B$.
Consequently, the $D_{l \mp 1, l \pm 1}^{(m,m')}(\Theta_B,
\phi_B)$ are complex functions, with imaginary part
%%%
\begin{eqnarray}
-\frac{i}{2} \left\{ D_{l \mp 1, l \pm 1}^{(m m')}- D_{l \mp 1, l
\pm 1}^{(m,m')\star} \right\} &=& {\mbox{Im}}\left(\langle a_{l
\mp 1, m}^\star a_{l \pm 1, m'}\rangle\right) \label{result+star2}
\\ \nonumber
&=& - \sin\Theta_B \sin\phi_B \Bigl\{ 2 \cos\Theta_B \
\Bigl[\delta_{m,m'-1} I_{\pm 2, -1}(m) - \delta_{m,m'+1} I_{\pm
2,-1}(-m)\Bigr]
\\ \nonumber
&&~~+\sin\Theta_B \cos\phi_B \Bigl[\delta_{m,m'-2}I_{\pm 2,
-2}(m) - \delta_{m,m'+2} I_{\pm 2, -2}(-m)\Bigr]\Bigr\}
I_d^{(l\mp 1,l \pm 1)} \pkt
\end{eqnarray}
%%%
So non-zero correlations between non-equal $m$ multipole numbers
result in an imaginary (antisymmetric) part of $\langle a^\star_{l
\mp 1,m} a_{l\pm 1, m'}\rangle$ which effectively breaks the
symmetry between the north and south hemispheres.

\section{Temperature correlations}

In this Section we derive the CMB temperature fluctuation
two-point correlation function $\langle {\Delta T}/{T}({\bf n})
{\Delta T}/{T}({\bf n'})\rangle$ that is induced purely from the
homogeneous magnetic field. This  must be added to the usual CMB
temperature two-point correlation function. Since the Alfv\'en
velocity is small this magnetic field induced CMB anisotropy is a
small correction to the ``primary'' CMB temperature fluctuations.
On the other hand, the effects that we discuss here  vanish in the
standard cosmological model so a non-zero correlation between $l$
and $l\pm 2$ multipole coefficients might indicate the presence
of an homogeneous cosmological magnetic field.

The two-point temperature correlation function can be written as
%%%
\begin{eqnarray} \label{deltaT}
\nonumber {\Big \langle} \frac{\Delta
T}{T}(\mathbf{n})\frac{\Delta T}{T}(\mathbf{n'}) {\Big \rangle}
&=& \frac{1}{2}\sum_{l, l'} \sum_{m,m'} \Bigl[ \langle
a_{lm}^\star a_{l' m'} \rangle Y_{lm}^\star(\mathbf{n}) Y_{l'
m'}(\mathbf{n'})+ \langle a_{lm} a^\star_{l', m'} \rangle Y_{l m}
(\mathbf{n}) Y^\star_{l' m'}(\mathbf{n'})\Bigr]  \nonumber \\&=&
{\Big \langle} \frac{\Delta T}{T}(\mathbf{n})\frac{\Delta
T}{T}(\mathbf{n'}) {\Big \rangle} {\Big |}^{l=l'} + {\Big \langle}
\frac{\Delta T}{T}(\mathbf{n})\frac{\Delta T}{T}(\mathbf{n'})
{\Big \rangle} {\Big |}^{l=l'\pm 2} \kma
\end{eqnarray}
%%%
where we introduce  complex conjugation to symmetrize over
$\mathbf{n}$ and $\mathbf{n'}$. From Eqs.~(\ref{30}) and
(\ref{1}) we see that both terms in the correlation function (the
contribution that are diagonal in $l$ as well as those that are
between $l$ and $l\pm 2$ multipole coefficients) contain three
kinds of terms, those proportional to: (1)  $3\cos^2\Theta_B-1$;
(2) $\sin\Theta_B \cos\Theta_Be^{\pm i\phi_B}$; and, (3)
$\sin^2\Theta_B e^{\pm 2 i\phi_B}$. We derive the contributions
from these terms in App.\ B.

Using the results of Sec.\ III for the multipole coefficients, and
the addition theorem of  Eq.~(\ref{vector-sum4}), we find, from
App.\ B,  the diagonal $l=l'$ correlation contribution,
%%%
\begin{eqnarray}
&&{\Big \langle} \frac{\Delta T}{T}(\mathbf{n}) \frac{\Delta
T}{T}(\mathbf{n'}) {\Big \rangle}{\Big |}^{l=l'} =
\frac{1}{4\pi}\sum_l \frac{l(l+1) (2l+1)}{(2l-1) (2l+3)} \Bigl\{
(2 l^2 +2 l-3)P_l
\label{llb}
\\
&&~~~~~~~~~+2(l^2 + l-3)  \Bigl[(\mathbf{b \cdot n}) (\mathbf{b
\cdot n'}) [2 P_{l-1}^{\prime\prime} +(2l-1) P_l^\prime ]
-[(\mathbf{b \cdot n})^2 +(\mathbf{b \cdot n'})^2 ]
P_l^{\prime\prime}+ P_{l-1}^\prime-l^2 P_l \Bigr]  \Bigr\}
I_d^{(l,l)}, \nonumber
\end{eqnarray}
%%%
and for the off-diagonal $l=l'\pm 2$ correlation contribution, where we use
the addition theorem of Eq.~(\ref{vector-sum3}), we find, from
App.\ B,
\begin{eqnarray}
&&\!\!\!\!\!\!\! {\Big \langle} \frac{\Delta T}{T}(\mathbf{n})
\frac{\Delta T}{T}(\mathbf{n'}) {\Big \rangle} {\Big |}^{l=l' \pm
2} =
\label{DTDKY2}
\\
&& ~~~~~~~\frac{1}{4\pi}\sum_l\frac{2(l+2)(l-1)}{2l+1}
\Bigl\{2(\mathbf{b\cdot n})(\mathbf{b \cdot n'})P_l^{''}
-\frac{1}{2} \Bigl[(\mathbf{b \cdot n})^2 + (\mathbf{b \cdot
n'})^2\Bigr] \Bigl[3P_l^\prime (x) +2 ({\bf n} \cdot {\bf n'})
P_l^{\prime \prime}\Bigr] + P_l^\prime\Bigr\}I_d^{(l-1,l+1)},
\nonumber
\end{eqnarray}
where the argument of the Legendre  polynomials and derivatives
in Eqs.~(\ref{llb}) and (\ref{DTDKY2})
are ${\bf n} \cdot {\bf n'}$.  If $\mathbf{b}$ is perpendicular
to $ \mathbf{n}$ or $\mathbf{n'}$, or if ${\bf n} = {\bf n'}$,
the above expressions simplify considerably.

To obtain the CMB temperature anisotropy two-point correlation
function, Eqs.~(\ref{llb}) and (\ref{DTDKY2}),  in terms of the
initial vorticity spectrum $P_{\Omega_0}= P_0
k^{n_\Omega}/k_S^{n_\Omega +3}$, the integrals $I_d^{(l,l)}$ and
$I_d^{(l-1,l+1)}$, Eq.~(\ref{int1}), must be evaluated. These can
be evaluated using an analytical approximations, for details see
App.\ A.3 and the Appendix of Ref.~\cite{kr05}.  The result
depends sensitively on the initial vorticity perturbation
spectral index ($n_\Omega$), Eq.~(\ref{spectrum}).

Accounting for the solution of Eq.~(\ref{vorticity-solution1}),
the symmetric part $P_\Omega$ of the resulting vorticity
perturbation spectrum is characterized by the spectral index
$n_\Omega+2$, i.e., $P_\Omega \propto k^{n_\Omega+2}$, while the
perturbed magnetic field ${\bf B_1}$ inherits the initial
vorticity spectral index $n_\Omega $. To avoid a divergence of
the energy density spectrum $E_\Omega$ of the resulting vorticity
perturbations on super-Hubble-radius scales, we require $n_\Omega
\geq -7$ ($E_\Omega (k) \propto k^{(n_\Omega +4)}$ and the
three-dimensional wavenumber integration gives an additional
factor of ${k}^3$). Requiring a non-divergent temperature
two-point correlation function at large wavenumbers leads to
$n_\Omega \leq -1$ \cite{dky98}.
%%% and the integrals $I_d^{(l l')}$ must be done according Eq.~(\ref{eq:GR-6.574.2}).
Another important value of $n_\Omega$ follows from the
requirement that the initial vorticity field energy density not
diverge at small wavenumbers, which results in $n_\Omega \geq
-5$. Requiring that the inequality $|{\bf\Omega}_0|^2 k^3 \leq
v_A^2$  (resulting from $B_1 \leq B_0$) \cite{dky98} hold on any
scale inside the Hubble radius at decoupling, i.e., for $k \geq
1/t_{\rm dec}$, we need \cite{dky98}
%%%
\begin{equation} \label{limits}
2P_0 \left(\frac{k}{k_S}\right)^{n_\Omega +3} \leq v_A^2 \kma
\end{equation}
%%%
which implies (accounting for $k \leq k_S$) $2P_0 \leq v_A^2$ for
$n_\Omega \geq -3$. As shown in Ref. \cite{dky98} this inequality
leads to an unconstrained magnetic field for $n_\Omega \geq -3$.
Since the more interesting results are  in the range $n_\Omega
\in (-7, -3)$, we adopt here $-3$ as the upper value for
$n_\Omega$. In this range of the spectral index $n_\Omega$ the
integral can be accurately computed analytically. When $n_\Omega
\geq -1$ the integral can be computed reasonable accurately in
the analytic approximation \cite{kr05}.

Using Eq.~(\ref{eq:GR-6.574.2}), for $n_\Omega \in (-7, -1)$, we
find,
\begin{eqnarray}
I_d^{(l,l)}&=& \frac{P_0~ v_A^2 ~\eta_{\rm dec}^2
~\Gamma(-n_\Omega/2-1/2) }{2\sqrt{\pi}~(k_S \eta_0)^{n_\Omega +3}
~\eta_0^2 ~\Gamma(-n_\Omega/2)}
~\frac{\Gamma(l+3/2+n_\Omega/2)}{\Gamma(l+1/2-n_\Omega/2)},
%l^{n_\Omega +1} \label{Idll}
\\
I_d^{(l-1,l+1)}&=& \frac{P_0~ v_A^2~ \eta_{\rm dec}^2
~(n_\Omega+2)\Gamma(-n_\Omega/2-1/2) }{2\sqrt{\pi}~(k_S
\eta_0)^{n_\Omega +3}~ \eta_0^2~ n_\Omega
\Gamma(-n_\Omega/2)}~\frac{\Gamma(l+3/2+n_\Omega/2)}
{\Gamma(l+1/2-n_\Omega/2)} \pkt
%l^{n_\Omega +1} \label{Idl-1l+1}
\end{eqnarray}
When $n_\Omega \geq -7$ the quadropole ($l=2$) moment does not
diverge, see the last term on the right hand sides $\propto
\Gamma(l+3/2+n_\Omega/2)$. For large enough $l$'s this last term
is $\propto l^{n_\Omega +1} $ and makes both integrals decay (for
$n_\Omega \le -1$) with $l$ as $l^{n_\Omega+1}$ for increasing
$l$.

\section{Conclusions}

We derive the CMB temperature anisotropy two-point correlation
function sourced by  vorticity perturbations induced by an
homogeneous magnetic field. We extend the analysis of Ref.\ \cite{dky98}
by considering a magnetic field that is
arbitrarily oriented with  respect to the  galactic  plane. We
consider a weak magnetic field, and since it is uniform and
points in a fixed direction it breaks spatial isotropy. In this case
the only non-zero correlations between multipole coefficients are
between those that have $\Delta l =0$ and $\Delta l=\pm 2$, and
$\Delta m=0$, $\Delta m = \pm 1$, and $\Delta m =\pm 2$, and we
have accounted for all non-zero correlations. Even though we have
computed only the two-point correlation function,  such
off-diagonal correlations indicate that in this model the CMB
temperature anisotropy is non-gaussian \cite{brown}. Such an
homogeneous magnetic field might explain the tentative large-scale
non-gaussianity of the CMB temperature anisotropy (also see
Refs.\ \cite{{cmb-anomalies-explanation-old},B,naselsky0}). Our
results, when used in analyses of the WMAP data, as well as
anticipated PLANCK data, could be used to search for or limit an
homogeneous cosmological magnetic field. The off-diagonal
correlations we have found might be a unique signature of such a
field.

While our results were obtained assuming an homogeneous magnetic
field, they can be extended to  an almost homogeneous
cosmological magnetic field with correlation length larger than
the Hubble radius today. Such a field, with a large enough
amplitude, can be generated by quantum-mechanical zero-point
fluctuations during inflation. In this case the spectral index of
the magnetic field is around $n_B = -3$. See Ref.\ \cite{ratra} and
the more recent studies in Refs.\ \cite{g}. Limits on a
cosmological magnetic field  that can be obtained through the
formalism we have developed here will compliment those obtained through
the CMB polarization Faraday rotation effect
\cite{kos,far,faraday,kmk08} and the non-zero cross-correlations
between CMB temperature and $B$-polarization anisotropies \cite{harari,ferreira}.

\acknowledgments
We greatly appreciate useful comments
and suggestions from P.~Naselsky. It is pleasure to thank L.~Samushia
for valuable remarks. We acknowledge helpful discussions
with R.~Durrer, M.~Hindmarsh, A.~Kosowsky,  and T.~Vachaspati.
T.~K. and G.~L. acknowledge the hospitality of the Abdus Salam
International Center for Theoretical Physics,  where part of this
work was  done, and  support from INTAS grant 061000017-9258.
T.~K. acknowledges support from Georgian NSF grant ST06/4-096.
T.~K. and B.~R. acknowledge support from DOE grant DE-FG03-99EP41093.

\begin{appendix}

\section{Useful Mathematical Formulae}

In this Appendix we list various mathematical results we use in
the computations.

\subsection{Spherical harmonics and Legendre polynomials}

The orthonormality relation for spherical harmonics is
\begin{equation}
\int d\Omega_{\bf\hat k} Y^\star_{rq}({\bf\hat k})Y_{r^\prime
q^\prime} ({\bf\hat k})=\delta_{rr^\prime}\delta_{qq^\prime} \pkt
\label{YY1}
\end{equation}

The recurrence relations for spherical harmonics are
\cite{arfken70}
\begin{eqnarray}
\cos\theta Y_{lm}(\theta,
\phi)&=&\alpha_{l+1,m}^{(0)}Y_{l+1,m}(\theta,\phi)
+\beta_{l-1,m}^{(0)}Y_{l-1,m}(\theta,\phi) \kma \label{cos-Y}
\\
\sin\theta e^{\pm i\phi } Y_{lm}(\theta, \phi) & = &
\alpha_{l+1,m \pm 1}^{(\pm)}Y_{l+1,m \pm 1}(\theta,\phi)
+\beta_{l-1,m \pm 1}^{(\pm)}Y_{l-1,m \pm 1}(\theta,\phi) \kma
\label{sin-Y}
\end{eqnarray}
where
\begin{eqnarray}
\alpha_{l,m}^{(0)} &=& \sqrt{\frac{(l-m)(l+m)}{(2l-1)(2l+1)}} \kma
~~~~~~~~~~~~~~~~~~~~~ \alpha_{l,m}^{(\pm)} = \mp
\sqrt{\frac{(l\pm m -1)(l \pm m)}{(2l-1)(2l+1)}} \kma
\label{alpha-def}
\\
\beta_{l,m}^{(0)} &=& \sqrt{\frac{(l-m+1)(l+m+1)}{(2l+1)(2l+3)}}
\kma ~~~~~~~~~~~\beta_{l,m}^{(\pm)} = \pm \sqrt{\frac{(l \mp m
+2)(l \mp m +1)}{(2l+1)(2l+3)}} \pkt \label{beta-def}
\end{eqnarray}

Legendre polynomials of order $l$ are defined by the sum
%\cite{jackson75}:
\begin{equation} \label{legendre-Y}
P_l({\bf n} \cdot {\bf n^\prime})=\frac{4\pi}{2l+1}\sum_{m=-l}^{l}
Y^\star_{lm}({\bf n})Y_{lm}({\bf n^\prime}) \pkt
%%% ~~~~~ P_1({\bf n}\cdot {\bf n^\prime})=\frac{4\pi}{3}\sum_{m=-1}^{1}
%%% Y^\star_{1m}({\bf n})Y_{1m}({\bf n^\prime}) \pkt
\end{equation}
Equations (\ref{YY1}) and (\ref{legendre-Y}) imply
%%%
\begin{equation}
\int d\Omega_{\bf\hat q} P_i({\bf n} \cdot {\bf \hat q}) P_j({\bf
n^\prime} \cdot {\bf \hat q}) = \frac{4\pi}{2j+1} \delta_{ij}
P_j({\bf n} \cdot {\bf n^\prime}) \pkt \label{legendre-ort}
\end{equation}
%%%

\subsection{Vector spherical harmonics} \label{vector_harmonics}

\subsubsection{Vector spherical harmonics components}

The $\mathbf{Y}_{lm}^{(\lambda)} (\mathbf{n}) (\lambda=-1, 0,+1)$
vector spherical harmonics are  (Eqs.~(6,7), p.~210,
\cite{varshalovich89})
\begin{eqnarray}
\mathbf{Y}_{lm}^{(+1)}(\mathbf{n})&=&\frac{1}{\sqrt{l(l+1)}}
\mathbf{\nabla}_\Omega
Y_{lm}(\mathbf{n}), \nonumber\\
\mathbf{Y}_{lm}^{(0)} (\mathbf{n})&=&\frac{-i}{\sqrt{l(l+1)}}
[\mathbf{n} \times
\mathbf{\nabla}_\Omega] Y_{lm}(\mathbf{n}), \nonumber\\
\mathbf{Y}_{lm}^{(-1)}(\mathbf{n})&=&\mathbf{n}
Y_{lm}(\mathbf{n}), \label{vectorspherical}
\end{eqnarray}
where $\mathbf{\nabla}_\Omega$ denotes the angular part of the
$\bf \nabla$ operator. The $\mathbf{Y}_{lm}^{(\lambda)}
(\mathbf{n})$ vector spherical harmonics are related to the ${\bf
Y}^{j}_{lm}({\bf n})$ vector spherical harmonics through
(Eq.~(9), p.~210, \cite{varshalovich89})
%%%
\begin{equation} \label{vector-lambda-def}
{\bf Y}_{rq}^{(\pm 1)}({\bf n}) = \sqrt{\frac{r}{2r+1}}{\bf Y}^{r
\pm 1}_{rq}({\bf n}) \pm \sqrt{\frac{r\pm 1}{2r+1}}{\bf Y}^{r \mp
1}_{rq}({\bf n}) \kma ~~~~~~~ {\bf Y}_{rq}^{(0)}({\bf n}) = {\bf
Y}_{rq}^r \pkt
\end{equation}
%%%

The ${\bf Y}^{j}_{lm}({\bf n})$ vector spherical harmonics are
related to the usual $Y_{lm}({\bf n})$ spherical harmonics through
(Eqs.~(9), (11-13), pp.~210-211, \cite{varshalovich89})
\begin{eqnarray}
&&{\bf n} Y_{lm}({\bf n})=
 \sqrt{\frac{l}{2l+1}}{\bf Y}^{l-1}_{lm}({\bf n})
- \sqrt{\frac{l+1}{2l+1}}{\bf Y}^{l+1}_{lm}({\bf n}) \kma
\label{vector-harmonics}
\\
&& {\bf Y}^{j}_{lm}({\bf n})=\sum_{s=-1}^1 |{\bf Y}^{j}_{lm}({\bf
n})|^s {\bf e}_s= \sum_{s=-1}^1 (-1)^s |{\bf Y}^{j}_{lm}({\bf
n})|_{-s} {\bf e}_s \kma
\\
\label{vector-harmonics2} && |{\bf Y}^{j}_{lm}({\bf n})|^s=
(-1)^s |{\bf Y}^{j}_{lm}({\bf n})|_{-s}=C_{j,m-s,1,s}^{lm}
Y_{j,m-s}({\bf n}) \kma \label{vector-harmonics-property}
\end{eqnarray}
where $ |{\bf Y}^{j}_{lm}({\bf n})|^s$ and $ |{\bf
Y}^{j}_{lm}({\bf n})|_s$ are contravariant and covariant
components, ${\bf e}_s (s=\pm 1,0)$ are unit covariant vectors,
and $C_{j,m-s,l,s}^{lm}$ are Clebsch-Gordon coefficients related
to the $\alpha_{l,m}^{(\pm)}$ and $\beta_{l,m}^{(\pm)}$
coefficients of  Eqs.~(\ref{alpha-def})--(\ref{beta-def}).

The contravariant components of the ${\bf Y}^{j}_{lm}({\bf n})$
vector spherical harmonics  are related to the usual spherical
harmonics through (pp.~211-212, \cite{varshalovich89})
%%%
\begin{eqnarray} \label{vector-components}
&&|{\bf Y}_{rq}^{r+1}({\bf n})|^{(\pm 1)} =\sqrt{\frac{(r \mp q
+1)(r \mp q +2)}{2(r+1)(2r+3)}} {Y}_{r+1,q \mp 1}({\bf n}) \kma ~~
|{\bf Y}_{rq}^{r+1}({\bf n})|^{(0)} =-\sqrt{\frac{(r - q +1)(r +q
+1)}{(r+1)(2r+3)}} {Y}_{r+1,q}({\bf n}) \kma
\nonumber \\
&&|{\bf Y}_{rq}^r({\bf n})|^{(\pm 1)} = \mp \sqrt{\frac{(r \pm
q)(r \mp q +1)}{2r(r+1)}} {Y}_{r, q \mp 1}({\bf n}) \kma
~~~~~~~~~~|{\bf Y}_{rq}^{r}({\bf n})|^{(0)}
=\frac{q}{\sqrt{r(r+1)}} {Y}_{r,q}({\bf n}) \kma
\nonumber \\
&&|{\bf Y}_{rq}^{r-1}({\bf n})|^{(\pm 1)} =\sqrt{\frac{(r \pm
q)(r \pm q -1)}{2r(2r-1)}} {Y}_{r-1,q \mp 1}({\bf n}) \kma
~~~~~~~|{\bf Y}_{rq}^{r-1}({\bf n})|^{(0)} =\sqrt{\frac{(r - q)(r
+q)}{r(2r-1)}} {Y}_{r-1,q}({\bf n}) \pkt
\end{eqnarray}
%%%

\subsubsection{Vector plane wave expansion}

A vector plane wave field can be expanded in vector spherical
harmonics (Eq.~(132), p.~228, \cite{varshalovich89}) as
\begin{equation}
{\bf v} (\mathbf{k}) e^{i\mathbf{k \cdot n}t}= \sum_{l,\lambda,m}
A_{lm}^{(\lambda)} \mathbf{Y}_{lm}^{(\lambda)}(\mathbf{n}),
\label{vdecom}
\end{equation}
where $\lambda=-1,0,1$, and the expansion coefficients for a
transverse field $\mathbf{v}(\mathbf{k})$
($\mathbf{v}(\mathbf{k})\cdot \mathbf{k}=0$)  are
%%%
\begin{eqnarray} \label{a-11}
A^{(-1)}_{lm}&=&4\pi i^{l-1} \sqrt{l(l+1)} \frac{j_{l}(kt)}{kt}
\mathbf{v}(\mathbf{k}) \cdot {\mathbf Y}_{lm}^{(+1)
\star}(\mathbf{\hat k}) \kma
\\ \label{a0}
A^{(0)}_{lm}&=&4\pi i^{l} j_{l}(kt) \mathbf{v}(\mathbf{k}) \cdot
{\mathbf Y}_{lm}^{(0) \star}(\mathbf{\hat k}) \kma
\\ \label{a+1}
A^{(+1)}_{lm}&=&-4\pi i^{l+1}\left(\frac{j_l(kt)}{kt} +
j_l^\prime(kt)\right) \mathbf{v}(\mathbf{k}) \cdot {\mathbf
Y}_{lm}^{(+1) \star}(\mathbf{\hat k}) \pkt
\end{eqnarray}
%%%
The terms  $\propto \mathbf{v}(\mathbf{k}) \cdot {\mathbf
Y}_{lm}^{(-1) \star}(\mathbf{\hat k})$ in the $A^{(\pm 1)}_{lm}$
coefficients vanish because $\mathbf{v}(\mathbf{\hat k}) \cdot
{\mathbf Y}_{lm}^{(-1) \star}(\mathbf{\hat k})=0$ as a
consequence of $\mathbf{\hat k}\cdot \mathbf{v}(\mathbf{k})=0$.

\subsubsection{Decomposition of vector spherical harmonics}

In the helicity basis where the angles $\Theta$ and $\phi$ are
defined by the unit wavevector  $\mathbf{\hat k}$, vector
spherical harmonics are given by (Eq.~(35), p.~215,
\cite{varshalovich89})

\begin{eqnarray} \label{y9}
\mathbf{Y}_{lm}^{(+1)}(\mathbf{\hat k})&=&\sqrt{\frac{2l+1}{8\pi}}
\left[D^{l}_{-1,-m}(0,\Theta, \phi) \mathbf{e'}_{+1}
(\mathbf{\hat k}) + D^{l}_{1,-m}(0,\Theta, \phi) \mathbf{e'}_{-1}
(\mathbf{\hat k}) \right],
\nonumber\\
\mathbf{Y}_{lm}^{(0)}(\mathbf{\hat k})&=&\sqrt{\frac{2l+1}{8\pi}}
\left[-D^{l}_{-1,-m}(0,\Theta, \phi) \mathbf{e'}_{+1}
(\mathbf{\hat k}) +D^{l}_{1,-m}(0,\Theta, \phi) \mathbf{e'}_{-1}
(\mathbf{\hat k}) \right],
\nonumber\\
\mathbf{Y}_{lm}^{(-1)}(\mathbf{\hat k})&=&\sqrt{\frac{2l+1}{4\pi}}
D^{l}_{0,-m}(0,\Theta, \phi) \mathbf{e'}_{0} (\mathbf{\hat k}).
\end{eqnarray}
Here the helicity basis vectors  ${\bf e}^{'}_{\mu}$ are defined
above Eq.\ (\ref{basis4}) and the Wigner $D$ functions are defined
(Eq.~(1), p.~76, \cite{varshalovich89}) by
\begin{equation} \label{w}
D^{l}_{m,m'}(\alpha,\beta, \gamma)=e^{-im\alpha}d^{l}_{mm'}(\beta)
e^{-im'\gamma} \kma
\end{equation}
where $d^l_{m,m'}(\beta)$ is a real function defined in Sec.~4.3
of Ref.~\cite{varshalovich89}.

\subsubsection{Addition theorems for and sums of vector spherical harmonics}

We have need for the following sums of  vector spherical harmonics
${\bf Y}_{rq}^{(\lambda)}({\bf n})$ (Eqs.~(80), p.~221,
\cite{varshalovich89})
%%%
\begin{equation} \label{sum-vec-1}
4\pi\sum_{q=-r}^{r}{Y}^\star_{rq}({\bf n})  {\bf
Y}_{rq}^{(-1)}({\bf n})=(2r+1){\bf n} \kma ~~~~~~
4\pi\sum_{q=-r}^{r}{Y}_{rq}^{\star}({\bf n})   {\bf
Y}_{rq}^{(0)}({\bf n}) =4\pi\sum_{q=-r}^{r}{Y}_{rq}^{\star}({\bf
n})   {\bf Y}_{rq}^{(1)}({\bf n})=0.
\end{equation}
%%%

Some addition theorems  for ${\bf Y}_{rq}^{R}$ are (p.~223,
\cite{varshalovich89}),
%%%
\begin{equation} \label{sum-vec-2}
4\pi\sum_{q=-r}^{r}{\bf Y}_{rq}^{R \star}({\bf n_1}) \cdot  {\bf
Y}_{rq}^{R^\prime}({\bf n_2})= \delta_{RR^\prime}(2r+1)P_R({\bf
n_1}\cdot {\bf n_2}) \kma
\end{equation}
%%%
and
%%%
\begin{equation} \label{sum-vec-3}
4\pi\sum_{q=-r}^{r}{\bf Y}_{rq}^{r\pm 1 \star}({\bf n_1})
\mathbf{\times}  {\bf Y}_{rq}^{r\mp 1}({\bf n_2})=0 \pkt
\end{equation}
%%%
The most general form of the addition theorems for vector
spherical harmonics are  given in Sect.\ 7.3.11 of
Ref.~\cite{varshalovich89}. Here we list two for arbitrary real
vectors ${\bf a}_1$ and ${\bf a}_2$.
%%%
\begin{eqnarray}
\label{vector-sum3} && 4\pi\sum_{q=-r}^{r} ({\bf a_1} \cdot  {\bf
Y}_{rq}^{r\pm 1\star}({\bf n_1}))   ({\bf a_2} \cdot {\bf
Y}_{rq}^{r \mp 1}({\bf n_2}))=
\frac{1}{\sqrt{r(r+1)}}\Bigl\{[({\bf a_1}\cdot {\bf n_1}) ({\bf
a_2}\cdot {\bf n_2}) + ({\bf a_1}\cdot {\bf n_2})({\bf a_2}\cdot
{\bf n_1})]P_r^{\prime \prime}  \Bigr.
\\ \nonumber
&&~~~~\qquad - \Bigl. ({\bf a_1}\cdot {\bf n_1}) ({\bf a_2}\cdot
{\bf n_1}) P_{r \pm 1}^{\prime \prime} -({\bf a_1}\cdot {\bf
n_2})({\bf a_2}\cdot {\bf n_2}) P_{r \mp 1}^{\prime \prime} +
({\bf a_1}\cdot {\bf a_2}) P_r^{\prime} \Bigr\} \kma
\\ \label{vector-sum4}
 &&4\pi\sum_{q=-r}^{r}({\bf a_1} \cdot  {\bf
Y}_{rq}^{r\star}({\bf n_1}))   ({\bf a_2} \cdot {\bf
Y}_{rq}^{r}({\bf n_2}))= \frac{2 r + 1}{r(r+1)} \Bigl\{-({\bf
a_1} \cdot {\bf n_1})({\bf a_2}\cdot {\bf n_2}) [P_{r-1}^{\prime\prime}
+ (r-1) P_r^{\prime}] \Bigr.
\\ \nonumber
&&~~~~\qquad -({\bf a_1}\cdot {\bf n_2})({\bf a_2}\cdot {\bf n_1})
[P_{r-1}^{\prime\prime} + r P_r^{\prime}] + [({\bf a_1}\cdot {\bf
n_1})({\bf a_2}\cdot {\bf n_1}) + ({\bf a_1}\cdot {\bf n_2})({\bf
a_2}\cdot {\bf n_2}) ] P_r^{\prime\prime} +({\bf a_1}\cdot {\bf
a_2}) [r^2 P_r - P_{r-1}^{\prime\prime}] \Bigl. \Bigr\} \pkt
\end{eqnarray}
%%%
In these expressions $P_r^\prime$ and $P_r^{\prime\prime}$ are
derivatives of Legendre polynomials and we have omitted the
arguments of Legendre polynomials and their derivatives,
abbreviating $P_r ({\bf n}_1 \cdot {\bf n'}_2)$ as $P_r$, etc.

\subsection{Integrals of spherical Bessel functions}

Here we present an analytical approximate formula to compute the
integral $I_d^{(l,l')}$ of Eq.~(\ref{int1}). The integrals that
we need to evaluate are of the form $\int^{x_S}_0 d
x\,J_p(ax)J_q(ax)x^{-b}$,
 which contain products of Bessel functions. For
$b >0$ when the integral converges and is dominated by $x\ll x_S$,
the upper limit $x_S$ can be replaced by $\infty$ (with an
accuracy of a few percent for $b>1$, and 15--30 \% for $0<b<1$,
depending on the value of $p-q$). We can then use Eq.~(6.574.2)
of  Ref.~\cite{gradshteyn94},
%%%
\begin{equation} \label{eq:GR-6.574.2}
\int^{\infty}_0dx\,J_p(ax)J_q(ax)x^{-b}=
\frac{a^{b-1}\Gamma(b)\Gamma((p+q-b+1)/{2})}
{2^b\Gamma((-p+q+b+1)/2) \Gamma((p+q+b+1)/{2})
\Gamma((p-q+b+1)/{2})} \kma
\end{equation}
%%%
which is valid for $\text{Re}\,(p+q+1)>\text{Re}\,b>0$ and $a>0$.

\section{Computation of temperature correlation functions}

The diagonal and off-diagonal correlation parts of the temperature
anisotropy two-point correlation function of Eq.~(\ref{deltaT}) are
\begin{eqnarray}\label{deltaTdiag}
{\Big\langle} \frac{\Delta T}{T}(\mathbf{n})\frac{\Delta
T}{T}(\mathbf{n'}) {\Big\rangle} {\Big |}^{l=l'} =
\frac{1}{2}\sum_{l} \sum_{m,m'} \left[ C_l^{(m, m')} Y_{l,
m}^\star(\mathbf{n}) Y_{l, m'}(\mathbf{n'}) + C_l^{(m, m')\star}
Y_{l, m} (\mathbf{n}) Y^\star_{l, m'}(\mathbf{n'})\right] \kma
\end{eqnarray}
%%%
and
%%%
\begin{eqnarray}
{\Big\langle} \frac{\Delta T}{T}(\mathbf{n})\frac{\Delta
T}{T}(\mathbf{n'}) {\Big\rangle } {\Big |}^{l=l'\pm 2} =
\frac{1}{2}\sum_{l} \sum_{m, m'} \!& {\Big [}
&\!\!\!D_{l-1,l+1}^{(m,m')} Y_{l-1, m}^\star(\mathbf{n}) Y_{l+1,
m'}(\mathbf{n'})
+ D_{l+1,l-1}^{(m,m')} Y_{l+1, m}^\star(\mathbf{n}) Y_{l-1, m'}(\mathbf{n'})
\nonumber
\\
&&\!\!\!+\left.\!D_{l-1,l+1}^{(m,m')\star} Y_{l-1, m} (\mathbf{n})
Y^\star_{l+1, m'}(\mathbf{n'}) + D_{l+1,l-1}^{(m,m')\star}
Y_{l+1, m} (\mathbf{n}) Y^\star_{l-1, m'}(\mathbf{n'}) \right]\pkt
\label{B2}
\end{eqnarray}
%%%
In this Appendix we summarize the results of a computation of
these terms.

We first compute the diagonal $l=l'$ correlations
of Eq.~(\ref{deltaTdiag}). From Eq.~(\ref{30}) we see that there
are three different types of terms, which we now list.
The first type of term is the $l=l'$ and $m=m'$ correlation
proportional to $3\cos^2\Theta_B-1$ on the right hand side of
Eq.~(\ref{30}), which results in
\begin{eqnarray} \label{llmm}
{4\pi} {\Big\langle} \frac{\Delta T}{T}(\mathbf{n}) \frac{\Delta
T}{T}(\mathbf{n'}) {\Big\rangle}{\Big |}^{l=l'}_{m=m'}\!\!\!\!
&=& \sum_l \frac{l(l+1)}{(2 l -1) (2l+3)}  \Bigl\{ \Bigr. (2l +1)
[l(l+1)+(l^2+l-3) b^0 b^0]
P_l ({\bf n}\cdot {\bf n'})  \nonumber \\
&&~~~~~- {{8\pi} (l^2+l-3)} \sum_m (b^0 b^0 +b^+b^-)
|\mathbf{Y}^{l}_{lm}(\mathbf{n})|^{0\star}
|\mathbf{Y}^{l}_{lm}(\mathbf{n'})|^0 \Bigl. \Bigr\} I_d^{(l,l)}
\pkt
\end{eqnarray}
%%%
The second type of term is the $l=l'$ and $m=m'\pm 1$ correlation
proportional to  $\sin\Theta_B\cos\Theta_Be^{\pm i\phi_B}$ on the
right hand side of Eq.~(\ref{30}), which results in
%%%
\begin{eqnarray} \label{llmm1}
{\Big \langle} \frac{\Delta T}{T}(\mathbf{n}) \frac{\Delta
T}{T}(\mathbf{n'}) {\Big \rangle} {\Big |}^{l=l'}_{m'=m\pm 1}
\!\!\!\!\!\!\! &&=\sum_{l,m} \frac{2 l(l+1) (l^2+l-3)}{(2 l -1)
(2 l +3)}\Bigl\{ b^0b^+ \Bigl[
|\mathbf{Y}^{l}_{lm}(\mathbf{n})|^{0\star}
|\mathbf{Y}^{l}_{lm}(\mathbf{n'})|^-  +
|\mathbf{Y}^{l}_{lm}(\mathbf{n})|^{-}
|\mathbf{Y}^{l}_{lm}(\mathbf{n'})|^{0\star} \Bigr] \Bigr.
\nonumber
\\
&&~~~~~~~+ b^0 b^- \Bigl[
|\mathbf{Y}^{l}_{lm}(\mathbf{n})|^{0\star}
|\mathbf{Y}^{l}_{lm}(\mathbf{n'})|^+ +
|\mathbf{Y}^{l}_{lm}(\mathbf{n})|^{+}
|\mathbf{Y}^{l}_{lm}(\mathbf{n'})|^{0\star} \Bigr] \Bigl.
\Bigr\} ~I_d^{(l,l)} \kma
\end{eqnarray}
%%%
where we have used Eqs.~(\ref{vector-components}).
The third type of term is the $l=l'$ and $m=m'\pm 2$ correlation
proportional to $\sin^2\Theta_B e^{\pm 2 i\phi_B}$ on the right
side of Eq.~(\ref{30}), which results in\footnote{For symmetry
reasons we have used correlations evaluated for $m\mp 1 = m'\pm
1$, and not for $m=m'\pm 2$. Thus in the expressions in
 Eqs.~(\ref{Isymmetry})  we replace $m$ by $ m \pm 1$. Of
 course, this does not affect the final results;
it just makes the computations easier and more symmetric.}
%%%
\begin{eqnarray} \label{llmm2}
&&{\Big\langle } \frac{\Delta T}{T}(\mathbf{n}) \frac{\Delta
T}{T}(\mathbf{n'}) {\Big \rangle}{\Big |}^{l=l'}_{m'=m\pm
2}\!\!\! =
\\  \nonumber
&& ~~~~~~~~~\sum_{l,m} \frac{2 l (l+1) (l^2+l-3)}{(2 l -1) (2 l
+3)} \Bigl\{ \Bigr.  b^+b^+
|\mathbf{Y}^{l}_{lm}(\mathbf{n})|^{+\star}
|\mathbf{Y}^{l}_{lm}(\mathbf{n'})|^- +b^-b^-
|\mathbf{Y}^{l-1}_{lm}(\mathbf{n})|^{-\star}
|\mathbf{Y}^{l+1}_{lm}(\mathbf{n'})|^+ \Bigl. \Bigr\} I_d^{(l,l)}
\pkt
\end{eqnarray}
%%%
$I_d^{(l,l)}$ in all three of these equations is defined in
Eq.~(\ref{int11}). Combining the expressions in
Eqs.~(\ref{llmm})--(\ref{llmm2}), we obtain
%%%
\begin{eqnarray} \label{lla}
{4\pi} {\Big \langle }\frac{\Delta T}{T}(\mathbf{n}) \frac{\Delta
T}{T}(\mathbf{n'}) {\Big \rangle} {\Big |}^{l=l'}\!\!\!\!\!&=&
\sum_l \frac{l(l+1)}{ (2l-1) (2l+3)}\Bigl\{ \Bigr. (2l+1)(2 l^2
+2 l-3) P_l ({\bf n}\cdot {\bf n'})
\\ \nonumber
&&~~~~- {4\pi} (l^2+l-3)\sum_m  \Bigl[ (\mathbf{b}
\cdot{\mathbf{Y}}_{lm}^{l}(\mathbf{n}))^\star (\mathbf{b}
\cdot{\mathbf{Y}}_{lm}^{l}(\mathbf{n'})) + (\mathbf{b}
\cdot{\mathbf{Y}}_{lm}^{l}(\mathbf{n})) (\mathbf{b}
\cdot{\mathbf{Y}}_{lm}^{l}(\mathbf{n'}))^\star \Bigr] \Bigl.
\Bigr\} I_d^{(l, l)} \pkt
\end{eqnarray}
%%%

There are three types of off-diagonal terms in Eq.~(\ref{B2}) (See Eq.~(\ref{1})),
similar to the diagonal case classified just above.
The first type of term is the $l=l'\pm 2$ and $m=m'$ correlation
proportional to $3\cos^2\Theta_B -1$ on the right hand side of
Eq.~(\ref{1}), which results in
%%%
\begin{eqnarray} \label{11}
&&\!\!\!\!\! \!\!\!\!\!\!\!\!\!\!\!\!\!{\Big \langle} \frac{\Delta
T}{T}(\mathbf{n}) \frac{\Delta T}{T}(\mathbf{n'}) {\Big
\rangle}{\Big |}^{l=l' \pm 2}_{m=m'} = \sum_{l,m} \frac{(l+2)
(l-1)\sqrt{l (l+1)}}{2 (2 l +1)} (b^0 b^0 +b^+b^-) \Bigl\{ \Bigr.
|\mathbf{Y}^{l-1}_{lm}(\mathbf{n})|^{0\star}
|\mathbf{Y}^{l+1}_{lm}(\mathbf{n'})|^0
\\ \nonumber
&&~~~~~~~~~~~~~~~~~~~~~~~~ +
|\mathbf{Y}^{l+1}_{lm}(\mathbf{n})|^{0\star}
|\mathbf{Y}^{l-1}_{lm}(\mathbf{n'})|^0+
|\mathbf{Y}^{l-1}_{lm}(\mathbf{n})|^0
|\mathbf{Y}^{l+1}_{lm}(\mathbf{n'})|^{0\star}+
|\mathbf{Y}^{l+1}_{lm}(\mathbf{n})|^0
|\mathbf{Y}^{l-1}_{lm}(\mathbf{n'})|^{0\star} \Bigl. \Bigr\}
I_d^{(l-1,l+1)} \pkt
\end{eqnarray}
%%%
Here we have used the relations $ \mathbf{b} \cdot
\mathbf{b}^\star= b^0 b^{\star 0} + b^+ b^{\star +}  +
b^-b^{\star -}= (b^0)^2-2b^+b^-=1$, $(\mathbf{Y}^L_{lm})^\star=
(-1)^{L+l+m+1}\mathbf{Y}^L_{l,-m}$, and
$|(\mathbf{Y}^L_{lm})^\star|^\mu= |(\mathbf{Y}^L_{lm})|_\mu$.
The second type of term is the $l=l'\pm 2$ and $m=m'\pm 1$
correlation proportional to
$\sin\Theta_B \cos\Theta_Be^{\pm i\phi_B}$
on the right hand side of Eq.\ (\ref{1}), which results in
%%%
\begin{eqnarray}
&&\!\!\!\!\!\!\!\!\!{\Big \langle} \frac{\Delta T}{T}(\mathbf{n})
\frac{\Delta T}{T}(\mathbf{n'}) {\Big \rangle}{\Big |}^{l=l' \pm
2}_{m'=m\pm 1}\!\! = -\sum_{l,m} \frac{(l+2) (l-1)\sqrt{l
(l+1)}}{(2 l +1)} I_d^{(l-1,l+1)}\label{21}
\\
&&\!\!\!\!\! \times \Bigl\{ b^0b^+ \Bigl[
|\mathbf{Y}^{l-1}_{lm}(\mathbf{n})|^{0\star}
|\mathbf{Y}^{l+1}_{lm}(\mathbf{n'})|^- +
|\mathbf{Y}^{l+1}_{lm}(\mathbf{n})|^{0\star}
|\mathbf{Y}^{l-1}_{lm}(\mathbf{n'})|^- -
|\mathbf{Y}^{l-1}_{lm}(\mathbf{n})|^{0}
|\mathbf{Y}^{l+1}_{lm}(\mathbf{n'})|^{+ \star } -
|\mathbf{Y}^{l+1}_{lm}(\mathbf{n})|^{0}
|\mathbf{Y}^{l-1}_{lm}(\mathbf{n'})|^{+ \star } \Bigr]
 \nonumber\\
 &&\!\!\!\!\!\!
 ~~~+ b^0 b^-\Bigl[
|\mathbf{Y}^{l-1}_{lm}(\mathbf{n})|^{0\star}
|\mathbf{Y}^{l+1}_{lm}(\mathbf{n'})|^+
+
|\mathbf{Y}^{l+1}_{lm}(\mathbf{n})|^{0\star}
|\mathbf{Y}^{l-1}_{lm}(\mathbf{n'})|^+|\mathbf{Y}^{l-1}_{lm}(\mathbf{n})|^{0}
|\mathbf{Y}^{l+1}_{lm}(\mathbf{n'})|^{- \star }
-
|\mathbf{Y}^{l+1}_{lm}(\mathbf{n})|^{0}
|\mathbf{Y}^{l-1}_{lm}(\mathbf{n'})|^{- \star } \Bigr] \Bigr\}~
\pkt\nonumber
\end{eqnarray}
%%%
Here we used Eqs.~(\ref{vector-components}).
The third type of term is the $l=l\pm 2$ and $m=m'\pm 2$
correlation proportional to $\sin^2\Theta_B e^{\pm 2 i\phi_B}$,
on the right hand side of Eq.~(\ref{1}), which results in (see
footnote 6)
%%%
\begin{eqnarray}
&&\!\!\!\!\!\!\!\!{\Big \langle} \frac{\Delta T}{T}(\mathbf{n})
\frac{\Delta T}{T}(\mathbf{n'}) {\Big \rangle}{\Big |}^{l=l' \pm
2}_{m'=m\pm 2}\!\! = - \sum_{l,m} \frac{(l+2) (l-1)\sqrt{l
(l+1)}}{2 (2 l +1)}I_d^{(l-1,l+1)} \label{3}\\
&&\!\!\!\!\!\!\!\!\times\Bigl\{ b^+b^+\Bigl[ \Bigr.
|\mathbf{Y}^{l-1}_{lm}(\mathbf{n})|^{+\star}
|\mathbf{Y}^{l+1}_{lm}(\mathbf{n'})|^- +
|\mathbf{Y}^{l+1}_{lm}(\mathbf{n})|^{+\star}
|\mathbf{Y}^{l-1}_{lm}(\mathbf{n'})|^-
+|\mathbf{Y}^{l-1}_{lm}(\mathbf{n})|^+
|\mathbf{Y}^{l+1}_{lm}(\mathbf{n'})|^{-\star} +
|\mathbf{Y}^{l+1}_{lm}(\mathbf{n})|^+
|\mathbf{Y}^{l-1}_{lm}(\mathbf{n'})|^{-\star} \Bigl. \Bigr]
\nonumber\\
&&\!\!\!\!\!\!\!\!~~ +b^-b^-\Bigl[ \Bigr.
|\mathbf{Y}^{l-1}_{lm}(\mathbf{n})|^{-\star}
|\mathbf{Y}^{l+1}_{lm}(\mathbf{n'})|^+ +
|\mathbf{Y}^{l+1}_{lm}(\mathbf{n})|^{-\star}
|\mathbf{Y}^{l-1}_{lm}(\mathbf{n'})|^+ +
|\mathbf{Y}^{l-1}_{lm}(\mathbf{n})|^-
|\mathbf{Y}^{l+1}_{lm}(\mathbf{n'})|^{+\star} +
|\mathbf{Y}^{l+1}_{lm}(\mathbf{n})|^-
|\mathbf{Y}^{l-1}_{lm}(\mathbf{n'})|^{+\star} \Bigl.  \Bigr]
\Bigr\}. \nonumber\end{eqnarray}
%%%
In these expressions $I_d^{(l, l')}$ is given in Eq.\ (\ref{int1}).
Combing the expressions in Eqs.\ (\ref{11})--(\ref{3}) and taking
into account that $\langle {\Delta T}/{T}(\mathbf{n}){\Delta
T}/{T}(\mathbf{n'}) \rangle =\langle{\Delta
T}/{T}(\mathbf{n}){\Delta T}/{T}(\mathbf{n'})\rangle^\star$
we obtain
%%%
\begin{eqnarray}
&&\!\!\!\!\!{\Big \langle} \frac{\Delta T}{T}(\mathbf{n})
\frac{\Delta T}{T}(\mathbf{n'}) {\Big \rangle} {\Big |}^{l=l' \pm
2}\!\!\! =\sum_{l,m} \frac{(l+2)(l-1)\sqrt{l(l+1)}}{2
(2l+1)}I_d^{(l-1, l+1)} \Bigl\{ (\mathbf{b}\cdot
{\mathbf{Y}}_{lm}^{l-1}(\mathbf{n}))^\star (\mathbf{b}\cdot
{\mathbf{Y}}_{lm}^{l+1}(\mathbf{n'})) \label{DTDKY22}\\&&
~~~~~~~~~~~~~~~~~~~~~~~~+(\mathbf{b}
\cdot{\mathbf{Y}}_{lm}^{l+1}(\mathbf{n}))^\star (\mathbf{b}
\cdot{\mathbf{Y}}_{lm}^{l-1}(\mathbf{n'})) + (\mathbf{b}
\cdot{\mathbf{Y}}_{lm}^{l-1}(\mathbf{n})) (\mathbf{b}\cdot
{\mathbf{Y}}_{lm}^{l+1}(\mathbf{n'}))^\star + (\mathbf{b}\cdot
{\mathbf{Y}}_{lm}^{l+1}(\mathbf{n})) (\mathbf{b}\cdot
{\mathbf{Y}}_{lm}^{l-1}(\mathbf{n'}))^\star \Bigr\}.\nonumber
\end{eqnarray}
%%%
\end{appendix}

%%%
\end{document}